\documentclass[preprint]{aastex}
\usepackage{amsbsy} 

\newcommand{\Msun}{M_\odot}
\newcommand{\bb}{\boldsymbol{b}} 
\newcommand{\bn}{\boldsymbol{n}} 
\newcommand{\by}{\boldsymbol{y}}
\newcommand{\bmu}{\boldsymbol{\mu}} 
\newcommand{\bnu}{\boldsymbol{\nu}}

\slugcomment{To Be Published in The Astrophysical Journal}

\shorttitle{Stellar Mass in the Pleiades}
\shortauthors{Converse \& Stahler}

\begin{document}

\title{The Distribution of Stellar Mass in the Pleiades}

\author{Joseph M. Converse and Steven W. Stahler}
\affil{Astronomy Department, University of California,
    Berkeley, CA 94720}

\email{jconverse@astro.berkeley.edu}

\begin{abstract}
As part of an effort to understand the origin of open clusters, we present a
statistical analysis of the currently observed Pleiades. Starting with a 
photometric catalog of the cluster, we employ a maximum likelihood technique
to determine the mass distribution of its members, including single 
stars and both components of binary systems. We find that the overall 
binary fraction for unresolved pairs is 68\%. Extrapolating to include 
resolved systems, this fraction climbs to about 76\%, significantly higher
than the accepted field-star result. Both figures are sensitive to the
cluster age, for which we have used the currently favored value of 125~Myr. 
The primary and secondary masses within binaries are correlated, in the sense 
that their ratios are closer to unity than under the hypothesis of random 
pairing. We map out the spatial variation of the cluster's projected and 
three-dimensional mass and number densities. Finally, we revisit the issue of 
mass segregation in the Pleiades. We find unambiguous evidence of segregation,
and introduce a new method for quantifying it.\end{abstract}

\keywords{open clusters and associations: individual (Pleiades) --- stars:
mass function, statistics --- binaries: general}

\section{Introduction}

Open clusters, with their dense central concentrations of stars, are relatively
easy to identify. Over a thousand systems are known, and the census is thought
to be complete out to 2~kpc \citep{br01,d02}. Because the clusters are no longer 
buried within interstellar gas and dust, their internal structure and dynamics
is also more accessible than for younger groups. 

Despite these favorable circumstances, many basic questions remain unanswered.
Most fundamentally, how do open clusters form? All observed systems have
undergone some degree of dynamical relaxation. Thus, the present-day 
distribution of stellar mass differs from the one just after disruption of the
parent cloud. Recovering this initial configuration will clearly be of value in 
addressing the formation issue. But such reconstruction presupposes, and 
indeed requires, that we gauge accurately the stellar content of 
{\it present-day} clusters. Even this simpler issue is a non-trivial one, 
as we show here.

We consider one of the most intensively studied open clusters, the Pleiades.
Again, our ultimate goal is to trace its evolution from the earliest, 
embedded phase to the present epoch. Focusing here on the latter, we ask 
the following questions: What are the actual masses of the member stars? 
Do they follow the field-star initial mass function? How many of the 
members are single, and how many are in binary pairs? Are the primary and 
secondary masses of binaries correlated? What is the overall density 
distribution in the cluster? What is the evidence for mass segregation, and
how can this phenomenon be quantified?

All of these questions have been addressed previously by others. 
\citet{dh04} constructed a global mass function for the Pleiades. Their method
was to assign masses based on the observed distribution of $R$-magnitudes. A 
more accurate assessment should account for the photometric influence of 
binaries. Several studies directed specifically at binaries have probed 
selected regions for spectroscopic pairs \citep{rm98,b97}. However, the 
overall binary fraction has not been carefully assessed, despite some 
preliminary attempts \citep{sj95,b97,m03}.

A fuller investigation of these and the other issues raised requires 
statistical methods; these should prove generally useful in characterizing 
stellar populations. Section~2 describes our approach, which employs a 
regularized maximum likelihood technique \citep[e.g.][]{c98}. A similar 
method has been applied to other astronomical problems, including the 
reconstruction of cloud shapes \citep{t07} and the investigation of 
binarity within globular clusters \citep{rw91}. Our study is the first to
apply this versatile tool to young stellar groups. In doing so, we also 
relax many of the restrictive assumptions adopted by previous researchers.
Section~3 presents our derived mass function for the Pleiades, along with
our results for binarity. The density distribution, as well our 
quantification of mass segregation, are the topics of Section~4. Finally,
Section~5 summarizes our findings, critically reexamines the binarity issue, 
and indicates the next steps in this continuing study.

\section{Method of Solution}
\subsection{Stellar Mass Probability Function}

The basic problem is how to assign stellar masses to all the point-like
sources believed to be cluster members. In many cases, the source is 
actually a spatially unresolved binary pair. More rarely, it is a triple 
or higher-order system; for simplicity, we ignore this possibility. The 
available observational data consists of photometry in several wavebands. 
Given the inevitable, random error associated with each photometric 
measurement, it is not possible to identify a unique primary and secondary
mass for each source. Instead we adopt a statistical approach that finds the
relative probability for each source to contain specific combinations of
primary and secondary masses.

We introduce a stellar mass probability density, to be denoted 
$\Phi (m_p, m_s)$. This two-dimensional function is defined so that 
$\Phi(m_p,m_s)\,\Delta m_p\,\Delta m_s$ is the probability that a binary
system exists with primary mass (hereafter expressed in solar units) in the 
interval $m_p$ to $m_p + \Delta m_p$ and secondary mass from $m_s$ to 
$m_s + \Delta m_s$. Single stars are viewed here as binaries with 
\hbox{$m_s = 0$}. We normalize the function over the full mass range:
\begin{equation}
\int_{m_{\rm min}}^{m_{\rm max}}\!dm_p
\int_0^{m_p}\!dm_s\,\Phi(m_p,m_s)\,=\,1 \,\,.
\end{equation}
Note that we integrate the secondary mass $m_s$ only to $m_p$, its maximum
value. Furthermore, we set the lower limit of $m_s$ to 0, in order to account
for single stars. It is assumed that \hbox{$\Phi\,=\,0$} for 
\hbox{$0\,<\,m_s\,<m_{\rm min}$}. Here, the global minimum mass $m_{\rm min}$ 
is taken to be 0.08, the brown dwarf limit. We consider $m_{\rm max}$ 
to be the highest mass still on the main sequence. The age of the Pleiades has 
been established from lithium dating as $1.25\times 10^8$~yr \citep{sta98}. 
This figure represents the main-sequence lifetime for a star of $4~\Msun$ 
\citep{s00}, which we adopt as $m_{\rm max}$. 

We examine separately the handful of stars that are ostensibly more massive 
than $m_{\rm max}$, and hence on post-main-sequence tracks. For these 11
sources, we assigned approximate masses from the observed spectral types, and 
obtained data on their known unresolved binary companions; this information 
was taken from the Bright Star Catalogue \citep{h91}. These systems were then 
added by hand to our mass functions. Finally, we ignore the brown dwarf 
population, thought to comprise from 10 to 16\% of the total system mass 
\citep{p98,sb05}.

The most direct method for evaluating $\Phi (m_p,m_s)$ would be to guess its
values over a discrete grid of $m_p$- and $m_s$-values. Each $(m_p,m_s)$ 
pair makes a certain contribution to the received flux in various 
wavebands. Thus, any guessed $\Phi (m_p,m_s)$ yields a predicted 
distribution of fluxes, which will generally differ from that observed. One
changes the guessed function until the observed flux distribution is most likely
to be a statistical realization of the predicted one.

Unfortunately, this straightforward approach is impractical. The basic 
difficulty is the mass-sensitivity of stellar luminosities. For secondary 
masses $m_s$ only modestly less than $m_p$, the binary is indistinguishable
photometrically from a single star having the primary mass. A $0.6\ \Msun$
main-sequence star, for example, has a $K$-band flux of 10.81~mag at the 
133~pc distance of the Pleiades \citep{s05}. Pairing this star with a
$0.2~\Msun$ secondary (which is not yet on the main sequence) only changes the
flux to 10.68~mag. In summary, the function $\Phi (m_p,m_s)$ evaluated in this
way is unconstrained throughout much of the $m_p - m_s$ plane.

\subsection{Correlation within Binaries}

Since binaries of even modestly large primary to secondary mass ratio are
difficult to recognize observationally, we need to infer their contribution
indirectly, within the context of a larger theoretical framework. The physical
origin of binaries is far from settled \citep{zm01}. There is a growing 
consensus, however, that most pairs form together, perhaps within a single 
dense core. The accumulating observations of protobinaries, i.e., tight pairs 
of Class~0 or Class~I infrared sources, have bolstered this view 
\citep[e.g.,][]{h04}. 

If binaries indeed form simultaneously, but independently, within a single 
dense core, there is no reason to expect a strong correlation between the
component masses. (Such a correlation would be expected if the secondaries 
formed within the primaries' circumstellar disks, for example.) A credible 
starting hypothesis, then, is that each component mass is selected randomly 
from its own probability distribution. If the formation mechanism of each star 
is identical, then these distributions are also the same. That is, we postulate 
that the true binary contribution to $\Phi (m_p,m_s)$ is
$\phi (m_p)\,\phi (m_s)$, where the single-star probability density $\phi$ is
properly normalized:
\begin{equation}
\int_{m_{\rm min}}^{m_{\rm max}}\!\phi (m)\,dm \,=\,1 \,\,.
\end{equation}

Of course, not all sources are unresolved binaries. Let $b$ represent the 
fraction of sources that are. We suppose that this fraction is independent of
stellar mass, provided the mass in question can represent either the primary
or secondary of a pair. While this hypothesis is reasonable for low-mass 
stars, it surely fails for O- and early B-type stars, which have an especially
high multiplicity \citep[e.g.,][]{m98,gm01}. Such massive objects, however, 
are not present in clusters of the Pleiades age.

Accepting the assumption of a global binary fraction, we have a tentative
expression for the full stellar mass probability:
\begin{equation}
\Phi (m_p,m_s) \,=\, 2\,b\,\phi (m_p)\,\phi (m_s) \,+\, (1\,-\,b)\,
                  \phi (m_p)\,\delta (m_s) \,\,.
\end{equation}  
Here, the first term represents true binaries, and the second single stars of
mass $m_p$. The factor of 2 multiplying the first term is necessary because of
the restricted range of integration for $m_s$ in equation~(1). That is, this
integration effectively covers only half of the $m_p$-$m_s$ plane. On the other
hand, the normalization condition of equation~(2) applies to both the primary
and secondary star, and covers the full range of mass, from $m_{\rm min}$ to 
$m_{\rm max}$, for each component.

We shall see below that the strict random pairing hypothesis, as expressed in
equation~(3), does not yield the optimal match between the predicted and 
observed distribution of magnitudes. The match can be improved, in the 
statistical sense outlined previously, if one allows for a limited degree of 
correlation between the primary and secondary masses within binaries. In 
other words, there is an apparent tendency for more massive primaries to be 
accompanied by secondaries that have a greater mass than would result from 
random selection.

A simple way to quantify this effect is to consider the extreme cases. If
there were {\it perfect} correlation between primary and secondary masses,
then the contribution to $\Phi (m_p,m_s)$ from binaries would be
$b\,\phi (m_p)\,\delta (m_p - m_s)$. With no correlation at all, 
$\Phi (m_p,m_s)$ is given by equation~(3). We accordingly define a correlation
coefficient $c$, whose value lies between 0 and 1. Our final expression for
$\Phi (m_p, m_s)$ uses $c$ to define a weighted average of the two extreme
cases:
\begin{equation}
\Phi (m_p,m_s) \,=\, 2\,b\,(1-c)\,\phi (m_p)\,\phi (m_s) \,+\,
                     b\,c\,\phi (m_p)\,\delta (m_p - m_s) \,+\, 
                     (1-b)\,\phi (m_p)\,\delta (m_s) \,\,.
\end{equation}
Note that the last righthand term, representing the probability of the source
being a single star, is unaffected by the degree of mass correlation within
binaries.

\subsection{Maximum Likelihood Analysis}
\subsubsection{From Masses to Magnitudes}

Reconstructing the stellar mass probability $\Phi(m_p,m_s)$ requires that we
evaluate the constants $b$ and $c$, as well as the single-star probability
$\phi (m)$. To deal with this continuous function, we divide the full mass
range into discrete bins of width $\Delta m_i$. Integrating over each bin,
we find $y_i$, the probability of a star's mass being in that narrow interval:
\begin{equation}
y_i \,\equiv\,\int_{m_i}^{m_i + \Delta m_i}\!\phi (m)\,dm \,\,. 
\end{equation}
We symbolize the full array of $y_i$-values by the vector $\by$, and 
similarly denote other arrays below. Our task, then, is to find optimal values 
not only for $b$ and $c$, but also for all but one element of $\by$. The 
normalization of $\phi$ is now expressed by the constraint
\begin{equation}
\sum_i y_i \,=\, 1 \,\,,
\end{equation}
which sets the last $\by$-value.

For each choice of $b$, $c$, and $\by$, equation~(4) tells us the relative 
probability of binaries being at any location in the $m_p - m_s$ plane. After 
dividing the plane into discrete bins, each labeled by an index $\alpha$, we 
define $\mu_\alpha$ as the predicted number of systems associated with a small 
bin centered on an $(m_p,m_s)$ pair. If $\mu_{\rm tot}$ is the total number of 
systems, i.e., of unresolved sources in all magnitude bins, then our chosen 
$b$, $c$, and $y_i$-values yield the relative fractions $\bmu/\mu_{\rm tot}$.

As an example, consider a bin $\alpha$ in which $m_p$ and $m_s$ have different
values lying between $m_{\rm min}$ and $m_{\rm max}$. Then the system is an 
unequal-mass binary, for which equation~(4) gives
\begin{equation}
{\mu_\alpha\over{\mu_{\rm tot}}} \,=\, 2\,b\,(1-c)\,y_p\,y_s \,\,.
\end{equation}
Here, $y_p$ is the element of $\by$ corresponding to the selected $m_p$, while
$y_s$ is similarly associated with $m_s$. For a bin where \hbox{$m_p\,=\,m_s$}, 
the corresponding relation is
\begin{equation}
{\mu_\alpha\over{\mu_{\rm tot}}} \,=\, b\,(1-c)\,y_p^2 \,+\, b\,c\,y_p \,\,.
\end{equation}
Note the additional term accounting for correlated binaries. Note also that a
factor of 2 has been dropped from equation~(4), since we are integrating only
over that portion of the mass bin with \hbox{$m_s\,<\,m_p$}. Finally, if the 
system is a single star, so that \hbox{$m_s\,=\,0$}, we have
\begin{equation} 
{\mu_\alpha\over{\mu_{\rm tot}}} \,=\, (1-b)\,y_p \,\,.
\end{equation} 

Our observational data consists of a catalog of $n_{\rm tot}$ sources, each of
which has an apparent magnitude in at least two broadband filters. (In practice,
these will be the $I$- and $K$-bands; see below.) As before, we divide this
two-dimensional magnitude space into small bins. Our choice of $b$, $c$, and
$\by$ leads not only to a predicted distribution in mass space, but also in
magnitude space.

Let $\nu_\beta$ be the predicted number of sources in each magnitude bin, now 
labeled by the index $\beta$. Then we may write the transformation from the 
mass to the magnitude distribution as
\begin{equation}
\nu_\beta \,=\, \sum_\alpha {\cal R}_{\beta\alpha}\,\mu_\alpha \,\,,
\end{equation} 
which may be recast in the abbreviated form
\begin{equation}
\bnu \,=\, {\cal R}\,{\bmu} \,\,. 
\end{equation}
Here, $\cal R$ is the {\it response matrix}, whose elements 
${\cal R}_{\alpha\beta}$ give the probability that a source in a mass bin 
$\alpha$ is observed in a magnitude bin $\beta$. In detail, this probability 
utilizes a theoretical isochrone in the color-magnitude diagram (see 
Section~3.1). We must also account for random errors in the measured 
photometry. In other words, a given magnitude pair can have contributions from
a range of mass pairs. It is for this reason that each element of $\bnu$ 
involves a {\it sum} over all $\alpha$-values.  

We previously showed how to obtain the relative mass distribution
$\bmu/\mu_{\rm tot}$, not the actual $\bmu$ itself. However, it is the latter 
that we need for equation~(11). To find $\mu_{\rm tot}$, we sum equation~(10) 
over all $\beta$-values, and demand that this sum be $n_{\rm tot}$, the total
number of observed sources:
\begin{displaymath}
n_{\rm tot} \,=\, \sum_\beta \nu_\beta \\
           \,=\, \sum_\beta \sum_\alpha {\cal R}_{\beta\alpha}\,\mu_\alpha\,\,, 
\end{displaymath}
so that
\begin{equation}
n_{\rm tot} \,=\, \mu_{\rm tot}\,\sum_\beta\sum_\alpha
           {\cal R}_{\beta\alpha}\
           \left(\frac{\mu_\alpha}{\mu_{\rm tot}}\right) \,\,.
\end{equation}
In summary, choosing $b$, $c$, and $\by$ gives us $\bmu/\mu_{\rm tot}$ through
through equations~(7), (8) and (9). We then solve equation~(12) for 
$\mu_{\rm tot}$. Supplied with knowledge of $\bmu$, we finally use
equation~(11) to compute $\bnu$.

\subsubsection{Likelihood and Regularization}

Having chosen $b$, $c$, and $\by$, how do we adjust these so that the predicted 
and observed magnitude distributions best match? Our technical treatment here 
closely follows that in \citet[][Chapter 11]{c98}, but specialized to our 
particular application. Let $n_\beta$ be the number of sources actually 
observed in each two-dimensional magnitude bin. We first seek the probability 
that the full array $\bn$ is a statistical realization of the predicted $\bnu$.
The supposition is that each element $\nu_\beta$ represents the {\it average} 
number of sources in the appropriate bin. This average would be attained after
a large number of random realizations of the underlying distribution. If 
individual observed values follow a Poisson distribution about the mean, then 
the probability of observing $n_\beta$ sources is
\begin{equation}
P\left(n_\beta\right) \,=\, {{{\nu_\beta}^{n_\beta}\,
                                   {\rm e}^{-\nu_\beta}}\over{{n_\beta}!}}\,\,.
\end{equation}
This probability is highest when $n_\beta$ is close to $\nu_\beta$.

The likelihood function $L$ is the total probability of observing the 
full set of $n_\beta$-values:
\begin{equation}
L \,\equiv\,\prod_\beta \, P (n_\beta) \,\,. 
\end{equation}
We will find it more convenient to deal with a sum rather than a product. Thus,
we use
\begin{equation}
{\rm ln}\,L \,=\, \sum_\beta\,n_\beta\,{\rm ln}\,\nu_\beta\,-\,\nu_\beta
                   \,-\,{\rm ln}\left(n_\beta !\right) \,\,.
\end{equation}
The strategy is then to find, for a given $\bn$, that $\bnu$ which maximizes 
${\rm ln}\,L$. For this purpose, we may neglect the third term in the sum, 
which does not depend on $\bnu$. We thus maximize a slightly modified function:
\begin{equation}
{\rm ln}\,L^\prime \,\equiv\,
\sum_\beta\,n_\beta\,{\rm ln}\,\nu_\beta\,-\,\nu_\beta \,\,.
\end{equation}  

Since, for a given $\bn$, each $P\left(n_\beta\right)$ peaks at
\hbox{$\nu_\beta\,=\,n_\beta$}, maximizing ${\rm ln}\,{L}^\prime$ is equivalent 
to setting $\bnu$ equal to $\bn$ in equation~(11), and then inverting the 
response matrix to obtain $\bmu$. Such a direct inversion procedure typically 
yields a very noisy $\bmu$, including unphysical (negative) elements. The 
solution is to {\it regularize} our result by employing an entropy term $S$:
\begin{equation}
S \,\equiv\, -\sum_i\,y_i\,\,{\rm ln}\left(y_i\right) \,\,.
\end{equation} 
The function $S$ is largest when the elements of $\by$ are evenly spread out. 
Adding $S$ to ${\rm ln}\,L^\prime$ and maximizing the total guarantees 
that the $\by$-values are smoothly distributed, i.e., that $\phi (m)$ is also 
a smooth function. 

In practice, we also want to vary the relative weighting of $S$ and
${\rm ln}\,L^\prime$. We do this by defining a {\it regularization parameter} 
$\lambda$, and then maximizing the function $\Gamma$, where
\begin{equation}
\Gamma \,\equiv\, \lambda\,{\rm ln}\,L^\prime \,+\, {\cal S} \,\,.
\end{equation}
For any given value of $\lambda$, maximizing $\Gamma$ yields an acceptably
smooth $\by$ that reproduces well the observed data. For the optimal solution, 
we find that value of $\lambda$ which gives the best balance between smoothness 
of the derived $\phi (m)$ and accuracy of fit. We do this by considering 
another statistical measure, the bias.

\subsubsection{Minimizing the Bias}

Our observational dataset, $\bn$, is an imperfect representation of the unknown 
probability density $\phi (m)$ in two senses. As already noted, $\bn$ may be
regarded as only one particular realization of the underlying distribution. 
Even this single realization would directly reveal $\phi (m)$ (or, 
equivalently, $\by$) if the sample size were infinite, which of course it is 
not.

Imagine that there were other realizations of $\phi (m)$. For each, we employ
our maximum likelihood technique to obtain $\by$. Averaging each $y_i$ over 
many trials yields its expectation value, $E (y_i)$. However, because of the 
finite sample size, $E (y_i)$ does not necessarily converge to the true value, 
$y_i^{\rm true}$. Their difference is the bias, $b_i$:
\begin{equation}
b_i \,\equiv\, E (y_i) \,-\, y_i^{\rm true} \,\,.
\end{equation}

The values of the biases, collectively denoted $\bb$, reflect the sensitivity 
of the estimated $\by$ to changes in $\bn$. Following 
\citet[][Section 11.6]{c98}, we define a 
matrix $\cal C$ with elements
\begin{equation}
{\cal C}_{i\beta} \,\equiv\, {{\partial y_i}\over{\partial n_\beta}} \,\,.
\end{equation} 
The bias is then given by
\begin{equation}
\bb \,=\, {\cal C} \,(\bnu \,-\, \bn )\,\,.
\end{equation}
To evaluate the derivatives in $\cal C$, we consider variations of the function
$\Gamma$ about its maximum. In matrix notation,
\begin{equation}
{\cal C} \,=\, -{\cal A}^{-1} {\cal B} \,\,,
\end{equation}
where the matrix $\cal A$ has elements
\begin{equation}
{\cal A}_{ij} \,\equiv\,
{{\partial^2 \Gamma}\over{\partial y_i\,\partial y_j}}\,\,,
\end{equation}
and $\cal B$ is given by
\begin{equation}
{\cal B}_{i\beta}\,\equiv\,
{{\partial^2 \Gamma}\over{\partial y_i\,\partial n_\beta}}
\,\,.
\end{equation}

Since $\Gamma$ is a known function of both the $\by$ and $\bn$, the derivatives 
appearing in both $\cal A$ and $\cal B$ may be evaluated analytically. Another 
matrix that will be useful shortly is $\cal D$, whose elements
\begin{equation}
{\cal D}_{\alpha i} \,\equiv\, {{\partial\mu_\alpha}\over{\partial y_i}}
\end{equation}
are also known analytically from equations~(7)-(9).

To determine the regularization parameter $\lambda$ appearing in $\Gamma$, we
seek to minimize the biases. In practice, we consider the weighted sum of their
squared values:
\begin{equation}
\chi_b^2 \,\equiv\, \sum_i {b_i^2\over {\cal W}_{ii}} \,\,,
\end{equation} 
and vary $\lambda$ to reduce this quantity.\footnote{In practice, we require 
that $\chi_b^2$ be reduced to $N$, the number of free parameters in our fit. As 
noted by \citet[][Section 11.7]{c98}, the average $b_i$-value at this point 
is about equal to its standard deviation, and so is statistically 
indistinguishable from zero.} Here, the ${\cal W}_{ii}$ are diagonal elements 
of $\cal W$, the covariance of the biases. Recall that the elements of $\bb$ 
are here considered to be random variables that change with different 
realizations.

We find the covariance matrix $\cal W$ by repeated application of the rule for
error propagation \citep[][Section 1.6]{c98}. We begin with $\cal V$, the
covariance of $\bn$. Since these values are assumed to be independently, 
Poisson-distributed variables, $\cal V$ has elements
\begin{equation} 
{\cal V}_{\alpha\beta} \,=\, \nu_\beta \,\delta_{\alpha\beta} \,\,. 
\end{equation} 
Here we have used the fact that the variance of the Poisson distribution 
equals its mean, $\nu_\beta$. In this equation only, both $\alpha$ and $\beta$
range over the $\nu$-values, i.e., $\cal V$ is a square matrix.

We next consider $\cal Y$, the covariance of $\by$. This is given by 
\begin{equation} 
{\cal Y} \,=\, {\cal C}\,{\cal V}\,{\cal C}^T \,\,. 
\end{equation}  
Finally, we obtain the desired $\cal W$ by
\begin{equation} 
{\cal W} \,=\, {\cal F}\,{\cal Y}\,{\cal F}^T \,\,.
\end{equation} 
The matrix $\cal F$ in this last equation has elements
\begin{equation}
{\cal F}_{ij} \,\equiv\, {{\partial b_i}\over{\partial y_j}} \,\,.
\end{equation}
Differentiating equation~(21) and applying the chain rule, we find $\cal F$ to
be
\begin{equation}
{\cal F} \,=\, {\cal C}\,{\bf\cal R}\,{\cal D}\,-\,{\cal I} \,\,,
\end{equation}
where $\cal I$ is the identity matrix.

\subsection{Calculation of Radial Structure}

Thus far, we have focused on determining global properties of the cluster,
especially the mass function $\phi (m)$. We also want to investigate the
spatial distribution of stellar masses. For this purpose, we need not perform
another maximum likelihood analysis. The reason is that we can treat the
mass distribution at each radius as a modification of the global result.

We divide the (projected) cluster into circular annuli, each centered on a 
radius $r$. What is $\mu^r_\alpha$, the number of sources in an annulus that
are within mass bin $\alpha$? (As before, each bin is labeled by the masses of
both binary components.) The quantity we seek is
\begin{equation}
\mu^r_\alpha \,=\, \sum_\beta {\cal Q}_{\alpha\beta}\,\nu_\beta^r \,\,.
\end{equation}
Here, ${\cal Q}_{\alpha\beta}$ is the probability that a source observed within 
magnitude bin $\beta$ has component masses within bin $\alpha$. In principle,
this probability depends on radius. For example, the source could have a high
probability of having a certain mass if there exist many such stars in that
annulus, even stars whose real magnitude is far from the observed one. If we 
discount such extreme variations in the underlying stellar population, then we
may approximate $\cal Q$ by its global average. 

The factor $\nu_\beta^r$ in equation~(32) is the estimated number of sources 
at radius $r$ in magnitude bin $\beta$. We only compute, via the maximum 
likelihood analysis, $\nu_\beta$, the estimated number of sources in the entire 
cluster. But we also know $n_\beta^r$, the {\it observed} number of sources in
the annulus. If the total number of observed sources, $n_\beta$, is non-zero, 
then we take
\begin{equation}
\nu_\beta^r \,\equiv\, \frac{n_\beta^r}{n_\beta}\,\nu_\beta \,\,. 
\end{equation}
In case \hbox{$n_\beta\,=\,0$}, then $n_\beta^r$ vanishes at all radii. We then
assume
\begin{equation}
\nu_\beta^r \,\equiv\, =\frac{n^r_{\rm tot}}{n_{\rm tot}}\,\nu_\beta \,\,,
\end{equation}
where $n^r_{\rm tot}$ is the observed source number of all magnitudes in the 
annulus. In the end, equation~(32) attributes the radial mass variation to 
changes in the local magnitude distribution, rather than to improbable 
observations of special objects.       

It is clear that the global $\cal Q$ must be closely related to the response 
matrix $\cal R$, which is the probability that a source with a given mass has a 
certain magnitude. The precise relation between the two follows from Bayes 
Theorem:
\begin{equation}
{\cal Q}_{\alpha\beta} \,=\, {\cal R}_{\beta\alpha}\,
\frac{\mu_\alpha/\mu_{\rm tot}}{\nu_\beta/\nu_{\rm tot}} \,\,.
\end{equation}
The numerator of the fraction is the probability that a source at any radius
lies within the mass bin $\alpha$, while the denominator is the probability of
it lying within magnitude bin $\beta$. Note that $\mu_{\rm tot}$ and 
$\nu_{\rm tot}$ need not be identical. The first quantity is the estimated 
number of sources covering all possible masses. The second is the observed 
number in the magnitude range under consideration. In practice, this range is
extensive enough that the two are nearly the same. We thus write
\begin{equation}
{\cal Q}_{\alpha\beta} \,=\, {\cal R}_{\beta\alpha}\, 
                          \frac{\mu_\alpha}{\nu_\beta}  \,\,.
\end{equation} 
Using this last equation, along with equation~(33), equation~(32) now becomes
\begin{equation}
\mu_\alpha^r \,=\, \mu_\alpha\,\sum_\beta {\cal R}_{\beta\alpha}\,
                \frac{n_\beta^r}{n_\beta} \,\,.
\end{equation}
For those terms where \hbox{$n_\beta\,=\,0$}, equation~(34) tells us to  
replace the ratio $n_\beta^r/n_\beta$ by $n_{\rm tot}^r/n_{\rm tot}$.

Summing $\mu_\alpha^r$ over all $\alpha$ and dividing by the area of the 
annulus gives the projected surface number density of sources as a function 
of radius. The total mass in each annulus is
\begin{equation}
\Delta m^r \,=\, \sum_\alpha \mu_\alpha^r\,m_\alpha \,\,,
\end{equation}
where $m_\alpha$ is the sum of the masses of both binary components in the 
appropriate bin. Division of $\Delta m^r$ by the annulus area gives the 
projected surface mass density. Under the assumption of spherical symmetry, 
the corresponding volume densities are then found by the standard 
transformation of the Abel integral \citep[][Section 4.2.3]{bm98}.  

\section{Application to the Pleiades: Global Results}

\subsection{The Response Matrix}

We begin with the observational data. Figure~1 is a dereddened $(I, I-K)$ 
color-magnitude diagram for the Pleiades, taken from the recent compilation by
\citet{sta07}. Shown are all sources which have high membership probability, as
gauged by their colors, radial velocities, and proper motions (see, e.g., 
\citet{dh04} for one such proper motion study.) The lower open circles 
correspond to probable brown dwarfs; we exclude such objects from our study.
Most brown dwarfs are too faint to be observed, and the population, in any case,
is more sparsely sampled. (The magnitude cutoffs corresponding to a 
$0.08~\Msun$ object are \hbox{$M_I\,=\,12$} and \hbox{$M_K\,=\,9$}.) After 
also exluding the 11 bright, post-main-sequence stars, shown here as large, 
filled circles, we have a total sample size of $n_{\rm tot} \,=\, 1245$.

The solid curve near the lower boundary of the stellar distribution is a 
combination of the theoretical zero-age main sequence for 
\hbox{$m_\ast \,>\, 1$} \citep{s00} and, for lower-mass stars, a
pre-main-sequence isochrone \citep{b98}.\footnote{Both theoretical results are
presented in magnitudes. We have applied corrections to the theoretical $K$-band
magnitudes to make them consistent with the 2MASS $K_s$-band used in Stauffer's
catalog. See \citet{c03} for this transformation.} The isochrone is that for 
the measured cluster age of 125~Myr. Our basic assumption is that the observed
scatter about this curve stems from two effects - binarity and intrinsic errors
in the photometric measurements. We do not consider, for now, possible 
uncertainty in the cluster's age. (See Section~5 for the effect of this 
uncertainty.) We also ignore the finite duration of star formation. This 
duration is roughly $10^7$~yr \citep{ps00}, or about 10\,\% of the cluster 
age.\footnote{We further ignore the effect of differential reddening across 
the cluster. \citet{sta07} adjusted individually the fluxes from sources in 
especially obscured regions, bringing their effective extinction to the 
observed average $A_V$ of 0.12. We therefore constructed Figure~1 by applying 
uniformly the corresponding $A_I$- and $A_K$-values of 0.06 and 0.01,
respectively.} 

After doing a polynomial fit to the mass-magnitude relations found by
\citet{s00} and \citet{b98}, we have analytic expressions for 
$M_I^\ast (m_p, m_s)$ and $M_K^\ast (m_p, m_s)$, the absolute $I$- and 
$K$-magnitudes for a binary consisting of a primary mass $m_p$ and
secondary $m_s$. Here, the superscripts indicate that the magnitudes are
theoretically derived. Both $M_I^\ast$ and $M_K^\ast$ are calculated by
appropriately combining the individual absolute magnitudes for $m_p$ and
$m_s$.

We do not actually observe $M_I^\ast$ or $M_K^\ast$ for any source. What we
have are dereddened, apparent magnitudes in these wavebands. Using the Pleiades
distance of 133~pc, these apparent magnitudes are readily converted to
absolute ones, $M_I$ and $M_K$. The salient question is: Given a source with
intrinsic magnitudes $M_I^\ast$ and $M_K^\ast$ (or, equivalently, with masses
$m_p$ and $m_s$), what is the probability that it is observed to have
magnitudes $M_I$ and $M_K$?

Here we confront the issue of photometric errors. We assume the errors in the
two wavebands to be normally distributed. Then the relevant probability
density is 
\begin{equation}
S\left(M_I,M_K;m_p,m_s\right) \,=\, \frac{1}{2\,\pi\,\sigma_I\,\sigma_K}\,
{\rm exp}\left[  
-\frac{\left(M_I\,-\,M_I^\ast\right)^2}{2\,\sigma_I^2}
-\frac{\left(M_K\,-\,M_K^\ast\right)^2}{2\,\sigma_K^2} 
\right]\,\,.
\end{equation}
Here, \hbox{$S\,\Delta M_I\,\Delta M_K$} is the probability of observing a
source in magnitude bin $\beta$, centered on \hbox{$(M_I, M_K)$}, and having
widths $\Delta M_I$ and $\Delta M_K$. 

The quantities $\sigma_I$ and $\sigma_K$ in equation~(39) are the standard
deviations of the photometric measurements. According to \citet{sta07}, the 
average standard deviation in the $I$-band is about 0.15. Figure~2, constructed
from Table~2 of \citet{sta07}, shows that $\sigma_K$ is generally lower, and 
rises steeply with $M_K$ for the dimmest sources.\footnote{This rise in 
$\sigma_K$ occurs because the observed $K$-magnitudes are approaching the 
sensitivity limit of the observations. Many of the $I$-band measurements 
come from POSS~II plates, for which the limit is 18.5 \citep{h93}. Another
large source of data was the observations of \citet{p00}, whose limiting
magnitude was 19.7. Our lower cutoff for brown dwarfs corresponds to an 
apparent $I$-magnitude of 17.7, so the rise in our $\sigma_I$ should be modest.}
The two branches of the curve presumably represent the results from two
different observations. We do a polynomial fit to the upper, majority, branch,
and thus have an explicit expression for $\sigma_K (M_K)$.

Suppose now that $m_p$ and $m_s$ are centered within a mass bin $\alpha$, which
has widths $\Delta m_p$  and $\Delta m_s$. Then the response matrix
${\cal R}_{\alpha\beta}$ is obtained by integrating $S (M_I,M_K;m_p,m_s)$ over
the magnitude bin, then averaging over the mass bin:
\begin{equation}
{\cal R}_{\alpha\beta} \,=\,\frac {
\int_{m_p}^{m_p + \Delta m_p}\!dm_p
\int_{m_s}^{m_s + \Delta m_s}\!dm_s
\int_{M_I}^{M_I + \Delta M_I}\!dM_I
\int_{M_K}^{M_K + \Delta M_K}\!dM_K
\,\,S} 
{\Delta m_p \,\Delta m_s} \,\,.
\end{equation}
The magnitude integrals can be expressed in terms of error functions if we 
reinterpret $\sigma_K$ as being a function of $M_K^\ast$ rather than $M_K$. 
The remaining numerical integrals over $m_p$ and $m_s$ are performed by 
finding, for each $(m_p,m_s)$ pair, the $M_I^\ast$- and $M_K^\ast$-values from 
our polynomial fits to the mass-magnitude relations.  

\subsection{Summary of Procedure and Synthetic Data Tests}

With the response matrix in hand, we are ready to input the catalog of source
magnitudes. Before turning to the Pleiades itself, we first employed a number
of synthetic datasets, in order to test various aspects of the code. We shall
describe these tests shortly. First, however, we summarize the standard
procedure we adopted for the analysis of any cluster, real or synthetic.

The basic problem, we recall, is to guess the optimal values of $b$, $c$, and
$\by$ that maximize the function $\Gamma$, as given in equation~(18). The 
entropy part of $\Gamma$, labeled $S$, is directly a function of $\by$ 
(eq.~(17)), while the modified likelihood function $L^\prime$ depends on the
observed magnitude distribution $\bn$ and the guessed one $\bnu$ (see eq.~(16).
The guessed $\by$ does not yield $\bnu$ itself, but the guessed mass 
distribution $\bmu$, through equations~(7)-(9). It is in this transformation 
that the binary fraction $b$ and correlation coefficient $c$ appear. Finally, 
$\bnu$ is obtained from $\bmu$ via the response matrix (eq.~(11)). 

We begin the maximization procedure by first setting the regularization
parameter $\lambda$ to zero. Since \hbox{$\Gamma\,=\,S$} in this case, the
optimal set of $\by$-values will be uniformly distributed, while subject to the
normalization constraint of equation~(6). We guess $b$, $c$, and $\by$, and 
vary them to maximize $\Gamma$. For the actual maximization, we employ a
standard simplex algorithm \citep[][Chapter 10]{p02}. The resulting 
best-fit parameters are then perturbed and the maximization rerun. This check, 
which may be redone several times, is done both to confirm convergence and to 
avoid becoming trapped in small, local maxima of the function $\Gamma$. We 
record the relevant covariances and biases, to be used in estimating errors in 
predicted quantities and to set the optimal $\lambda$-value. 

The next step is to increase $\lambda$ slightly. We maximize $\Gamma$ in the 
same way as before, again recording covariances and biases. We again increase 
$\lambda$, repeating the entire procedure until $\chi_b^2$, the weighted sum 
of the biases, starts to become acceptably small. At this point, the best-fit 
$b$, $c$, and $\by$ have been established.

As a first test of the procedure, we introduced an artificial cluster whose
single-star probability, $\phi (m)$, we selected beforehand. Sources were
chosen randomly to have masses according to this distribution. In a certain 
fraction of the sources, our preset binary fraction $b$, a second star was
added to the source. The mass of this object was also randomly chosen from
$\phi (m)$. Thus, the correlation coefficient $c$ was initially zero. Given
both masses in a source, its intrinsic $M_I^\ast$ and $M_K^\ast$ are readily
obtained. These magnitudes are smeared out into neighboring bins according to 
Gaussian errors with the appropriate standard deviations $\sigma_I$ and
$\sigma_K$. Thus, the ``observed" magnitude distribution, $\bn$, is 
established. 

Figure~3 shows two representative examples. In the left panel, the chosen
$\phi (m)$, is a power law: 
\hbox{$\phi (m) \,\propto\, m^{-2.8}$}. On the right, we used a log-normal
distribution:
\begin{equation}
\phi (m) \,=\,{C\over m}\,{\rm exp}  
\left[-\frac{({\rm log}\,m \,-\,{\rm log}\,m_\circ)^2}{2\,\sigma_m^2}
\right] \,\,,
\end{equation}
where $C$ is the normalization constant. The central mass was chosen as
\hbox{$m_\circ\,=\,0.2$} and the logarithmic width $\sigma_m$ was 0.4. The 
binary fraction $b$ was chosen to be 0.30 in the power-law example, and 0.68 
for the log-normal distribution. The total source number $n_{\rm tot}$ was 
10,000 in both cases. 

The smooth curve in both panels is $\phi(m)$, while the data points are the
best-fit values of $y_i/\Delta m_i$, where $\Delta m_i$ is the bin width. Shown
also is the estimated error for each value. This was derived from the 
covariance matrix $\cal Y$, introduced in Section~(2.1). Specifically, the 
plotted error is $\sqrt{{\cal Y}_{ii}}/\Delta m_i$. We divide each $y_i$ and 
its associated error by $\Delta m_i$ because $y_i$ is integrated over the bin 
(eq.~(5)).

It is evident that the code reproduces well the assumed $\phi (m)$ in these
two examples. Note that most of the scatter seen in both plots, especially in 
the left panel, was already present in the input data, which were finite 
realizations of the analytic distributions. The derived (i.e., predicted) 
binary fractions, 
\hbox{$b\,=\,0.293\pm 0.008$} and \hbox{$b\,=\,0.672\pm 0.011$}, respectively,
are also in good agreement. We had similar success when we reduced 
$n_{\rm tot}$ to 1245, the actual number in our Pleiades source catalog. In this
smaller sample, the errors in our predicted mass function and binary fraction 
increased, roughly as ${n_{\rm tot}}^{-1/2}$. 

Figure~4, taken from a dataset with \hbox{$n_{\rm tot}\,=\,1245$}, shows 
in more detail how the regularization parameter $\lambda$ was chosen. The 
figure also illustrates some of the subtlety involved in this procedure. 
Plotted here, as a function of $\lambda$, is $\chi_b^2$, defined in 
equation~(26). As $\lambda$ is gradually increased, $\chi_b^2$ takes a sudden,
sharp dip. After climbing back, $\chi_b^2$ then more slowly declines, 
eventually falling below \hbox{$N\,=\,21$}, the number of tunable parameters 
in this maximization ($b$, $c$, and 19 $\by$-values).

It is the second threshold (\hbox{$\lambda\,=\,0.027$} in this case) that marks
the true stopping point. The earlier dip in $\chi_b^2$ is due, not to a 
decrease in the biases, but to a sharp {\it increase} in the covariances 
$\cal W$. This increase commonly occurs when the likelihood term 
${\rm ln}\,L^\prime$ starts to become comparable to the entropy $S$ in 
the full function $\Gamma$. At that point, $\by$ makes an abrupt shift away 
from its earlier, nearly uniform, distribution. With further increase in 
$\lambda$, $\by$ settles down gradually to its optimal form.   

Continuing our synthetic data tests, we next introduced a correlation between 
the primary and secondary masses. First, we generated uncorrelated pairs, as 
above. Generalizing the prescription of \citet{d96}, we then altered the 
secondary mass in each source according to 
\begin{equation}
m_s \,\rightarrow\, m_s \left(\frac{m_p}{m_s}\right)^\gamma \,\,.
\end{equation} 
Here, $\gamma$, a preset number between 0 and 1, represents our imposed degree
of correlation. Thus, setting \hbox{$\gamma\,=\,0$} yields the previous, 
uncorrelated case, while, for \hbox{$\gamma\,=\,1$}, every binary has 
equal-mass components. We ran our routine with a variety of input single-star 
mass functions, binary fractions, and degrees of correlation. 

Our general result was that the predicted $\by$ still reproduced well the
synthetic $\phi (m)$. The binary fraction $b$ was similarly accurate. Most
significantly, the predicted $c$-value tracked the input quantity $\gamma$.
Figure~5 shows this relation. We conclude that our statistical model, while
crudely accounting for correlation by inserting a fraction of equal-mass pairs,
nevertheless mimics a smoother correlation, such as would be found naturally. 
The shaded patch in the figure is the probable region occupied by the real 
Pleiades; Section~3.4 below justifies this assessment.\footnote{The 
prescription for mass correlation given in equation (42) is more realistic 
than introducing a subpopulation of identical-mass binaries (eq.~(4)). We 
employed the latter device only for convenience. If we had parametrized the 
correlation through $\gamma$, equations~(7) and (8) would have been numerical 
integrals, and the derivative matrix $\cal D$ in equation~(25) would also have required
numerical evaluation.} 

One price we paid for our simplified account of correlation was that our 
matching of $\phi (m)$ was less accurate than for randomly paired input 
binaries. Consider, for example, the log-normal function of equation~(41).
While our best-fit $\by$ still reproduced $\phi (m)$ reasonably well, the
output function peaked at too high a mass compared to $m_\circ$. The filled
circles in Figure~6 shows that this shift, $\Delta m_\circ$, increased with
the input $\gamma$-value. Concurrently, our output function was too narrow
compared to the input $\sigma_m$. The (negative) difference, $\Delta\sigma_m$,
displayed as open circles in Figure~6, was also more pronounced at higher
$\gamma$. These systematic errors need to be considered when analyzing a real
cluster. The two patched areas in the figure again represent the likely
regime of the Pleiades, as we explain shortly.

\subsection{Empirical Mass Distributions} 

We now present the results of applying our maximum likelihood analysis to the
Pleiades itself, i.e., to the $I$- and $K$-magnitudes of 1245 sources from the
catalog of \citet{sta07}. Our best-fit binary fraction was
\hbox{$b\,=\,0.68\,\pm\,0.02$}, while the correlation coefficient was
\hbox{$c\,=\,0.36\,\pm\,0.06$}. (These and other uncertainties represent only
random statistical error, and do not include systematic effects; see 
Section~5.) We will discuss the implications of these findings in the following
section. First, we examine the global distribution of stellar mass.

The data points in Figure~7 are the best-fit values of each $y_i/\Delta m_i$.
As in Figure~3, these points are a discrete representation of the
single-star mass function $\phi (m)$. The large error bars on the two points at
highest mass are due to the small number of sources gauged to be in the 
respective bins. The smooth, solid curve in Figure~7 is a log-normal mass 
function that best matches the empirical $\by$. Referring again to 
equation~(41), we find that \hbox{$m_\circ\,=\,0.20\,\pm\,0.04$} and 
\hbox{$\sigma_m\,=\,0.38\,\pm\,0.02$}. The presence of a finite binary 
correlation affects both estimates. Judging from Figure~6, our $m_\circ$ is 
overestimated by about 0.06, while $\sigma_m$ should be raised by 0.08.

Each of our mass bins has contributions from both the primary and 
secondary components of binary pairs, as quantified by equation~(4). Integrating
the full stellar mass probablity $\Phi (m_p,m_s)$ over all secondary masses,
we obtain $\phi_p (m_p)$, the probability distribution of primary masses:
\begin{equation}
\phi_p (m_p) \,=\, \int_0^{m_p}\! dm_s\,\Phi (m_p,m_s)  \,\,.
\end{equation}
Note that this distribution includes the possibility that the star is single
(\hbox{$m_s \,=\,0$}). 

The solid curve in Figure~8 is a log-normal fit to the empirical $\phi_p (m_p)$.
Shown for comparison as a dashed curve is the fit for $\phi (m)$ from Figure~7.
Relative to the latter function, the primary distribution falls off at lower 
masses. This falloff simply reflects the fact that less massive objects are 
more likely to be part of a binary containing a higher-mass star, and thus to 
be labeled as ``secondaries."\footnote{We may similarly calculate a secondary 
mass distribution $\phi_s (m_s)$ by integrating $\Phi (m_p,m_s)$ over $m_p$, 
from $m_s$ to $m_{\rm max}$. The function $\phi_s$ has an excess of low-mass 
stars and drops very steeply at high masses, as most such objects are 
primaries.} In any event, we now see why the peak of $\phi_p (m_p)$, 
\hbox{$m_\circ\,=\,0.27\,\pm\,0.02$}, is elevated with respect to the peak of
$\phi (m)$. Similarly, the primary distribution is also slightly narrower than
the single-star mass function, with \hbox{$\sigma_m\,=\,0.35\,\pm\,0.01$}. 

The parameters of our log-normal approximation to $\phi_p (m_p)$ may be 
compared to those of \citet{m04}. These authors fit the entire mass function.
Since, however, they did not account for binarity, their results are more
closely analogous to our primary distribution. Their best-fit $m_\circ$ of
0.25 is close to ours, while their $\sigma_m$ of 0.52 is higher, mostly
because of their inclusion of the highest-mass members. These parameters are
also close to those given by \citet{ch03} in his log-normal fit to the
field-star initial mass function (\hbox{$m_\circ\,=\,0.22$}, 
\hbox{$\sigma_m\,=\,0.57$}).

In Figure~9, we compare our single-star distribution $\phi (m)$ to the 
field-star initial mass function (dashed curve). The latter, which has been
raised in the figure for clarity, is taken from \citet{k01}, who did correct 
for binarity. It is apparent that $\phi (m)$ itself veers away from the IMF 
for both low- and high-mass objects. When these are added in, the resemblance 
improves. The open circles in Figure~9 are Pleiades low-mass stars and brown 
dwarfs found by \citet{bi06}. We have normalized their data, taken from a 
limited area of the cluster, so that their total number of stars matches ours 
within the overlapping mass range. No such normalization was necessary for the
11 B-type stars (filled circles), which are from the catalog of \citet{sta07} 
but not included in our maximum likelihood analysis. Adding both these groups 
not only improves the match to the IMF, but also reveals a gap in the stellar 
distribution between about 2 and $5\,\,\Msun$. A similar gap is seen in the 
Pleiades mass function of \citet[][see their Figure~1]{m04}. 

Our estimate for the total cluster mass, based solely on the 1245 catalog
sources, is $738\,\,\Msun$, with a 4\% uncertainty. Adding in the brightest
stars brings this total to $820\,\,\Msun$, with the same relative error. Tests
with synthetic data indicate that the systematic bias due to binary correlation 
raises this figure by roughly $50\,\,\Msun$, to $870\,\,\Msun$. Addition of the 
brown dwarfs would cause a further, relatively small, increase. For comparison, 
\citet{p98} found $735\,\,\Msun$ in stars, and an upper limit of $131\,\,\Msun$ 
for the brown dwarf contribution. \citet{rm98} used the virial theorem to 
estimate a total mass of $720\,\,\Msun$, with a 28\% uncertainty. Direct 
integration of their mass function gave $950\,\,\Msun$, with an 18\% fractional 
error.

\subsection{Binarity}

The global binary fraction, \hbox{$b\,=\,0.68$}, obtained in our analysis
represents most, but not all, of the full binary population. Omitted here are
spatially resolved systems. For these, the primary and secondary appear as
separate sources in the catalog of \citet{sta98}. Counting resolved pairs
raises the total fraction to about 76\%, as we now show.

The smallest angular separation between stars in the catalog is 
$10^{\prime\prime}$. At the Pleiades distance of 133~pc, the corresponding
physical separation is 1400~AU. An edge-on circular binary of exactly this
orbital diameter will still be unresolved, since the components spend most of
their time closer together. The true minimum separation in this case is
2200~AU. Here, we have divided 1400~AU by $2/\pi$, which is the average of
${\vert\rm sin}\,\theta\vert$, for $\theta$ randomly distributed between 0 and
$2\pi$.

Of course, only a relatively small fraction of binaries have separations
exceeding 2200~AU. The average total mass of our unresolved systems is 
$0.71\,\,\Msun$. A binary of that total mass and a 2200~AU diameter has a
period of $1.2\times 10^5\,\,{\rm yr}$. What fraction of binaries have even
longer periods? Our average {\it primary} mass is $0.46\,\,\Msun$, corresponding
to a spectral type of M1. \citet{fm92} studied the period distribution of 
binaries containing M-type primaries. They claimed that this distribution
was indistinguishable from that found by \citet{dm91} for G-type primaries. In
this latter sample, 11\% of the systems had periods greater than our limiting
value. If the Pleiades periods are similarly distributed, then the total
fraction of binaries - both resolved and unresolved - becomes
\hbox{$0.66/(1-0.11)\,=\,0.76$}. 

Even without this augmentation, our total binary fraction appears to be
inconsistent with the available direct observations of the Pleiades. Thus, 
\citet{b97} found visual pairs with periods between 40 and 
$3.4\times 10^4$~yr. Using the period distribution of \citet{dm91} to 
extrapolate their observed binary fraction of 28\% yields a total fraction of 
60\%. \citet{m92} observed spectroscopic pairs with periods under 3~yr. A 
similar exercise again yields 60\%. We note, however, that this ostensible 
concurrence of results is based on very broad extrapolations from limited data.
(See Fig.~4 of \citet{b97}.) 

Our derived binary fraction also exceeds that found in the field-star 
population. \citet{dm91} found that 57\% of G~stars are the primaries of
binary or higher-order systems. Note that our $b$ represents the total
probability that a star is in a binary, whether as the primary or secondary
component. Since G~stars are rarely secondaries, the comparison with
\citet{dm91} is appropriate. On the other hand, M~stars {\it are} frequently 
secondaries, so we would expect the fraction of binaries with M-type primaries  
to be reduced. \citet{l06} has found that only 25\% of M-stars are the
primary components of binaries. Our own analysis yields a binary fraction of 
45\% for M-star primaries, still in excess of the field-star result.

If our finding of a relatively high binary fraction proves robust, it may 
provide a clue to the progenitor state of the Pleiades and other open clusters.
A similar statement applies to the correlation between component masses within
binaries. Our adopted method of gauging this correlation - inserting a 
fraction of equal-mass pairs in the mass function - is admittedly crude. 
Nevertheless, the strong result (\hbox{$c\,=\,0.36\,\pm\,0.06$}) is 
significant. Referring back to Figure~5, we find that the Pleiades correlation
is equivalent to setting $\gamma$ equal to about 0.65 in the alternative 
description of equation~(42). Whatever the origin of the Pleiades binaries, 
the primaries and secondaries were not formed by completely independent 
processes.  

\section{Application to the Pleiades: Radial Distributions}

\subsection{Number and Mass Profiles}

We now employ the procedure outlined in Section~2.4 to investigate both the 
surface and volumetric density as a function of the projected distance from 
the cluster center. The filled circles in Figure~10, along with the associated
error bars, represent the surface number density of sources, measured in 
pc$^{-2}$. The solid curve is a density profile using the empirical 
prescription of \citet{k62}. Here, the core radius is 2.1~pc, while the tidal 
radius is 19~pc. For comparison, \citet{a01} also fit the surface number 
density profile of their low-mass stars to a King model, with a core radius of
2.3-3.0~pc. Our profile is also at least roughly consistent with the 
cumulative number distribution displayed by \citet{rm98}. Our best-fit tidal 
radius is slightly larger than the 17~pc cited by these authors.

The surface mass density is plotted in an analogous fashion, again as a function
of the projected radius. We show both the data points (small open circles) and, 
as the dashed curve, the best-fit King model. Here, the core radius is 
1.3~pc, and the tidal radius is 18~pc. Note that the mass density profile
falls off more steeply than the number density. Thus, stars near the center 
are abnormally massive, a trend we shall explore more extensively below.

Figure~11 displays the corresponding volumetric densities. As we indicated, the 
deconvolution from surface profiles assumes spherical symmetry. In fact, the
Pleiades is slightly asymmetric, with a projected axis ratio of 1.2:1
\citep{rm98}. This ellipticity is thought to stem from the tidal component of
the Galactic gravitational field \citep{w74}. Under the spherical assumption,
the filled circles and solid, smooth curve show the number density. Here,
the King model is the same used for the surface number density in Figure~10, 
but deprojected into three-dimensional space. 

Figure~11 also shows, as the small open circles and dashed curve, the mass 
density as a function of spherical radius. Again, the King model here is the
deprojected version of that from Figure~10. The relatively rapid falloff in
the mass, as opposed to the number, density is another sign of the tendency
for more massive stars to crowd toward the center.

The information we used in obtaining these profiles also gives us the spatial 
variation of the binary fraction $b$. That is, we first used equation~(37)
to obtain $\bmu^r$, the predicted mass distribution in each radial annulus. 
Recall that the distribution refers to both primaries and secondaries, 
as well as single stars. The binary fraction can thus be computed locally. To 
within our uncertainty, about $\pm 0.05$ at each radial bin, we find no 
variation of $b$ across the cluster.

\subsection{Mass Segregation}

We have mentioned, in a qualitative manner, that more massive cluster members
tend to reside nearer the center. In Figure~12, we explicitly show this trend.
Here, we plot $\langle m_p + m_s\rangle$, the average system mass (primary 
plus secondary) as a function of the projected cluster radius. It is apparent 
that $\langle m_p + m_s\rangle$ monotonically falls out to about 4~pc. Beyond 
that point, the average mass is roughly constant.

The pattern here is consistent with mass segregation, but is not a clear
demonstration of that effect. The problem is that Figure~12 gives no indication
of the relative populations at different annuli. If the outer ones are occupied
by only a small fraction of the cluster, is mass segregation present? To gauge
any variation in the mass distribution of stars, that distribution must be
calculated over an adequate sample size.

Previous authors have also claimed evidence of mass segregation, using various
criteria. \citet{a01} looked at the distribution of surface and volumetric 
number densities for a number of different mass bins. \citet{rm98} divided the
population by magnitude into relatively bright and faint stars. They calculated
the cumulative number as a function of radius for both groups, and found the
bright stars to be more centrally concentrated. Finally, \citet{p98} fit King
profiles to the surface density of various mass bins. As the average mass
increases, the core radius shrinks. 

Figure~13 gives a simpler and more clear-cut demonstration of the effect. Here,
we consider $f_N$, the number of sources enclused in a given projected radius,
divided by the total number of sources in the cluster. We also consider $f_M$,
the analogous fractional mass inside any projected radius. The figure then
plots $f_M$ versus $f_N$. In the absence of mass segregation, $f_M$ would 
equal $f_N$ at each annulus. This hypothetical situation is illustrated by the 
dotted diagonal line.

In reality, $f_M$ rises above $f_N$, before they both reach unity at the cluster
boundary (see upper smooth curve, along with the data points). This rise, as
already noted, indicates that the innermost portion of the cluster has an 
anomalously large average mass, i.e., that mass segregation is present. 
Moreover, the {\it area} between the solid and dotted curves is a direct 
measure of the effect. In the case of ``perfect" mass segregation, a few stars
near the center would contain virtually all the cluster mass. The solid curve 
would trace the upper rectangular boundary of the plot, and the enclosed area 
would be 0.5. We thus define the {\it Gini coefficient}, $G$, as {\it twice} 
the area between the actual $f_M - f_N$ curve and the central 
diagonal.\footnote{The name derives from economics, where the coefficient is 
used to measure inequality in the distribution of wealth \citep{s97}. As we 
show in the Appendix, $G$ is also half the mean mass difference of radial 
shells, where that mass difference is normalized to the average system mass in
the cluster.} For the Pleiades, we find that \hbox{$G \,=\,0.20\pm 0.06$}.  

It is possible, at least in principle, that this effect is due entirely to a few
exceptionally massive stars located near the center. In fact, this is {\it not}
the case. We have artificially removed the 11 brightest sources (all late-B 
stars) and recalculated $f_M$ versus $f_N$. The result is shown by the dashed 
curve in Figure~13. While the rise above the diagonal is diminished, it is 
still present. That is, the intermediate-mass population exhibits segregation,
as well.

An interesting contrast is presented by another populous group, the Orion
Nebula Cluster (ONC). The distribution of stellar masses in this far younger 
system was recently studied by \citet{hs06}. Figure~2 in that paper compares the
stellar populations in the inner and outer halves of the
cluster.\footnote{Note that the axis labels in Figure~2 of \citet{hs06} were
inadvertantly switched.} Apart from a few high-mass objects, the two 
populations are essential identical. 

We may also construct an $f_M - f_N$ curve for the ONC, as shown here in 
Figure~14. The solid curve again lies well above the fiducial diagonal, 
ostensibly indicating mass segregation. However, removal of just the four 
Trapezium stars gives a dramatically different result (dashed curve) that is 
virtually indistinquishable from the diagonal. All stars except this tiny subset
are similarly distributed. The cluster is too young to have undergone true mass 
segregation, a conclusion drawn previously from N-body simulations 
\citep{bd98}. The Trapezium represents a special population, one that probably
formed just prior to cloud dispersal \citep{hs07}.

\section{Discussion}

In this paper, we have applied a versatile statistical tool, the maximum
likelihood technique, to assess the distribution of stellar mass and the
incidence of binaries in the Pleiades. We began with a near-infrared catalog of
cluster members. Our basic assumption was that all cluster members share the
same evolutionary age, and that any dispersion in the color-magnitude plane
stems from binarity and random photometric errors. We were then able to infer
the most probable distribution of masses, both for the cluster as a whole,
and as a function of distance from its center. Finally, we introduced a simple 
method for gauging the degree of mass segregation in the cluster.

One of our surprising results is the relatively high fraction of binaries. We
estimate that 68\% of all systems in the cluster are unresolved binaries; this
figure climbs to about 76\% if resolved pairs are included. These fractions are
significantly higher than the accepted field-star result, so we should
scrutinize them carefully. Could they stem from an underestimate of the 
photometric error at faint magnitudes? Since the error in $I$ is greater than 
$K$, we artificially increased the dispersion $\sigma_I$. We kept $\sigma_I$ 
at 0.15 until \hbox{$M_I \,=\,9.5$}, below which we increased it linearly, 
reaching \hbox{$\sigma_I\,=\,0.20$} at \hbox{$M_I \,=\, 12$}. After redoing the 
maximum likelihood analysis, the global binary fraction $b$ for unresolved 
pairs is unchanged.

Another potential difficulty is our neglect of the physical thickness of the
cluster. We have assigned all members a distance of 133~pc, although there will
naturally be some variation. However, this effect is also relatively small.
From Section~4.1, the volumetric number density falls off with radius 
approximately as a King model with core and tidal radii of 2.1 and 19~pc,
respectively. Consider the front half of a spherical cluster with such a 
density distribution. It may readily be shown that the mean distance from the 
plane of the sky of any cluster member is \hbox{$d\,=\,2.6\,{\rm pc}$}. For a
cluster at mean distance $D$, the induced magnitude spread is 
\hbox{$5\ {\rm log}\,[(D+d)/D]$}, which is 0.04 in our case. Although the
actual spread in magnitudes is not Gaussian, we have added this figure in
quadrature to both $\sigma_I$ and $\sigma_K$, and rerun the analysis. Again,
the binarity is unaffected.

The errors due to both photometry and finite cluster thickness induce a 
symmetric spread in stellar magnitudes. That is, they scatter as many sources
below the fiducial isochrone as above it. Thus, they cannot reduce the estimated
binarity, which stems from an excess of stars above the isochrone. One 
systematic error that {\it would} affect $b$ is an overestimation of the 
cluster distance. If $D$ were lowered, the absolute magnitudes of all sources
would decrease equally, as would the inferred $b$-value. Quantitatively, the
distance would have to decrease by about 15~pc to bring the binary fraction 
down to the field-star result for G-dwarf primaries. An error of this size for 
the average distance is excluded by current observations, for which the 
estimated uncertainty is only 1~pc \citep{s05}.

Since our method relies solely on photometry to assess binarity, we cannot
distinguish between physically linked pairs and chance alignments. As mentioned
in Section~3.4, the resolution limit of our data is $10^{\prime\prime}$, or
\hbox{$\Delta r_\circ \,=\, 1400~{\rm AU}$} at the distance of the Pleiades.
Consider a star at a radius $r$ from the cluster center. Its average number   
of neighbors within $\Delta r_\circ$ is 
\hbox{$\pi\,\Delta r_\circ^2\,n_s (r)$},where $n_s (r)$ is the projected 
surface number density of the cluster. Since each ring of width $dr$ contains 
\hbox{$2\,\pi\,n_s\,r\,dr$} stars, integration over all members yields the 
total number of chance alignments:
\begin{equation}
N_{\rm chance} \,=\, 2\,\pi^2\,\Delta r_\circ^2 \int_0^R\!n_s^2 (r)\,r\,dr
\,\,. \end{equation}
where $R$ is the cluster's outer radius, Using $n_s (r)$ from Figure~10, we 
find\hbox{$N_{\rm chance} \,=\,2.4$}. Thus, chance alignments have no 
quantitative impact. 

Yet another source of systematic error is the cluster age. We have adopted the
lithium-based figure of 125~Myr from \citet{sta98}. Earlier estimates, using
the main-sequence turnoff, yielded a range of answers. For example, 
\citet{m93} found 100~Myr. Even this minor reduction affects our results, since
it lifts the low-mass end of the isochrone toward higher luminosity. For an 
age of 100~Myr, our analysis gives \hbox{$b\,=\,0.57\,\pm\,0.02$} and 
\hbox{$c\,=\,0.28\,\pm\,0.06$}. The binary fraction is augmented to 0.64 when we
include resolved pairs. From Figure~3 of \citet{sta98}, a 100~Myr age
corresponds to a lithium edge at \hbox{$M_I\,=\,11.7$}, or \hbox{$I\,=\,17.3$}
at the Pleiades distance. Such a result seems incompatible with the lithium 
data shown in Figure~2 of \citet{sta98}, but the total number of observations 
is relatively small.

We conclude that the enhanced binarity is a real effect. What this fact tells
us about the origin of open clusters remains to be seen. Our next step in
addressing this basic issue is to try and account theoretically for the
empirical properties just obtained through our statistical analysis. We will
ascertain, using direct numerical simulations, the range of initial states
that can relax dynamically to the present-day Pleiades. Such a study will
bring us one step closer to understanding the molecular cloud environments
that give rise to open clusters.  

\acknowledgments

We are grateful to Eric Huff, who first suggested the use of the maximum
likelihood technique for this problem and provided continued insight. We also
benefited from conversations with James Graham and Geoff Marcy. The referee's
comments helped to improve the final presentation. This research was supported 
by NSF grant~AST-0639743. 

\clearpage



\appendix

\section{Interpreting the Gini Coefficient}

In Section 4.2, we introduced the Gini coefficient geometrically, as twice the
area between the $f_M - f_N$ curve and the diagnonal line representing zero
mass segregation. Alternatively, $G$ may be defined in terms of the mean mass 
difference between radial shells in the cluster. Here we describe more 
precisely, and prove the equivalence of, this second interpretation. 

Altering our previous notation, we now let $m (r)$ be the average system mass
(primary plus secondary) in a shell with outer radius $r$. If there are many 
shells, then we may define a system number density $n (r)$, such that the
number of systems between $r$ and $r + dr$ is $n(r)\,dr$. The total number of 
systems at all radii is
\begin{equation}
N_{\rm tot} \,=\,\int_0^\infty\!n(r)\,dr \,\,.
\end{equation}
This total was called $\mu_{\rm tot}$ in the text. The average system mass
throughout the entire cluster is
\begin{equation}
{\bar m} \,=\, \frac{1}{N_{\rm tot}}
\int_0^\infty\! 
m(r)\,n(r)\,dr \,\,.
\end{equation}
We will be concerned with the {\it relative mean difference} in the mass of
shells. This is
\begin{equation}
{\bar\Delta} \,=\,\frac{1}{{\bar m}\,N_{\rm tot}^2}
\int_0^\infty\!dr
\int_0^\infty\!dr^\prime\,
{\vert m (r)\,-\,m (r^\prime)\vert}\,\,n (r)\,\,n (r^\prime) \,\,.
\end{equation}
We will prove that the Gini coefficient, as defined in Section~4.2, is also
$\bar\Delta /2$.

In the geometric definition of $G$, we utilized the cumulative fractional 
number $f_N$ and cumulative fractional mass $f_M$. These may be written in 
terms of the system number density:
\begin{eqnarray} 
f_N (r) & = & 
\frac{1}{N_{\rm tot}} 
\int_0^r\!n (r^\prime)\,dr^\prime \,\,, \\
f_M (r) & = & 
\frac{1}{{\bar m}\,N_{\rm tot}}
\int_0^r\!m (r^\prime)\,n (r^\prime)\,dr^\prime \,\,.
\end{eqnarray} 
We will later need the differentials of $f_N$ and $f_M$, which are
\begin{eqnarray}
{df}\!_N & = &
\frac{n (r)}{N_{\rm tot}}\,\,dr  \,\,, \\
{df}\!_M & = &
\frac{m (r)\,n (r)}{{\bar m}\,N_{\rm tot}}\,
\,\,dr \,\,.
\end{eqnarray} \
In terms of $f_N$ and $f_M$, the geometric definition of $G$ is then
\begin{equation}
G \,=\, 2\int_0^1\!\left(f_M \,-\,f_N \right)\,df_N \,\,.
\end{equation}
If the mass is centrally concentrated, as expected in a real stellar cluster, 
then \hbox{$f_M\,\ge\,f_N$} at all radii, and \hbox{$G\,\ge\,0$}. Hypothetical
clusters in which larger masses are preferentially located farther from the
center would have \hbox{$G\,\le\,0$}. 

We now return to equation~(A3) and manipulate it to obtain $G$, as defined
above. A central assumption we will make is that $m (r)$ declines monotonically
with $r$. We may then expand the righthand side of equation~(A3). We split 
the integral over $r^\prime$ into two parts, one for \hbox{$r^\prime \le r$}
and the other for \hbox{$r^\prime > r$}. Under our assumption, 
\hbox{$m (r^\prime)\,\ge\,m (r)$} in the first integral, and
\hbox{$m (r^\prime)\,<\,m (r)$} in the second. After pulling $n (r)$ from the
$r^\prime$-integration, we further split the integrands to find
\begin{equation}
{\bar\Delta} \,= \, \frac{1}{{\bar m}\,N_{\rm tot}}
\int_0^\infty\! 
(-{\cal I}_1 \,+\, {\cal I}_2 \,+\, {\cal I}_3 \,-\, {\cal I}_4)\,\,
n(r)\,dr \,\,.
\end{equation}
The first two terms of the integrand are
\begin{eqnarray}
{\cal I}_1 & \equiv & 
\frac{m (r)}{N_{\rm tot}}
\int_0^r\!n (r^\prime)\,dr^\prime \,=\, m (r)\,f_N (r) \,\,,\\
{\cal I}_2 & \equiv & 
\frac{1}{N_{\rm tot}}
\int_0^r\!m (r^\prime)\,n (r^\prime)\,dr^\prime \,=\, {\bar m}\,f_M (r) \,\,.
\end{eqnarray}
The third term is
\begin{eqnarray}
{\cal I}_3 & \equiv & 
\frac{m (r)}{N_{\rm tot}}
\int_r^\infty\!n (r^\prime)\,dr^\prime  \\
           & = & 
\frac{m (r)}{N_{\rm tot}}
\int_0^\infty\!n (r^\prime)\,dr^\prime \,-\, 
\frac{m (r)}{N_{\rm tot}}
\int_0^r\!n (r^\prime)\,dr^\prime \\
           & = &
m (r)\,-\, m (r)\,f_N (r) \,\,,
\end{eqnarray}
while the fourth is 
\begin{eqnarray}
{\cal I}_4 & \equiv &
\frac{1}{N_{\rm tot}}
\int_r^\infty\!m (r^\prime)\,n (r^\prime)\,dr^\prime  \\
           & = &
\frac{1}{N_{\rm tot}}
\int_0^\infty\!m (r^\prime)\,n (r^\prime)\,dr^\prime \,-\,
\frac{1}{N_{\rm tot}}
\int_0^r\!m (r^\prime)\,n (r^\prime)\,dr^\prime \\ 
	   & = & 
{\bar m} \,-\, {\bar m}\,f_M (r)\,\,. 
\end{eqnarray} 

Putting equations~(A10), (A11), (A14), and (A17) back into equation~(A9) 
yields
\begin{eqnarray}
\bar\Delta & = & 
\frac{1}{{\bar m}\,N_{\rm tot}}
\int_0^\infty\! 
\left[ 2\,{\bar m}\,f_M (r) \,-\, 2\,m (r)\,f_N (r)
\,+\, m (r) \,-\, {\bar m} \right]\,
n (r)\,dr \\
           & = &
2\,\left( \int_0^1\!f_M\,d f_N \,-\, \int_0^1\!f_N\,d f_M \right) \,\,,
\end{eqnarray}
where we have used both equations~(A1) and (A2) to cancel the last two terms
on the right side of equation~(A18), and equations (A6) and (A7) to transform
the remaining two terms. The second integral in the last equation is
\begin{eqnarray}
\int_0^1\!f_N\,d f_M & = & 
\int_{f_N f_M = 0}^{f_N f_M = 1}
\!d\,(f_N f_M) \,-\, \int_0^1\!f_M\,d f_N  \\
                     & = &
1 \,-\, \int_0^1\! f_M \,d f_N \,\,,
\end{eqnarray}
since $f_M$ and $f_N$ attain their upper and lower bounds simultaneously. Using 
this result, equation~(A19) becomes
\begin{equation}
{\bar\Delta} \,=\, 2\,\left( 2\int_0^1\!f_M\,d f_N \,-\,1\right) \,\,.
\end{equation}

Finally, we note that
\begin{equation}
2 \int_0^1\!f_N\,d f_N \, = \, 1 \,\,.
\end{equation}
Thus, equation~(A22) becomes
\begin{eqnarray}
{\bar\Delta} & = &
4\,\left( \int_0^1\!f_M\,d f_N \,-\, \int_0^1\!f_N\,d f_N \right) \\
             & = &
4\int_0^1\!\left( f_M\,-\,f_N \right)\,d f_N \\
             & = &
2\,G \,\,,             
\end{eqnarray}
as claimed.

\clearpage

\begin{figure}
\plotone{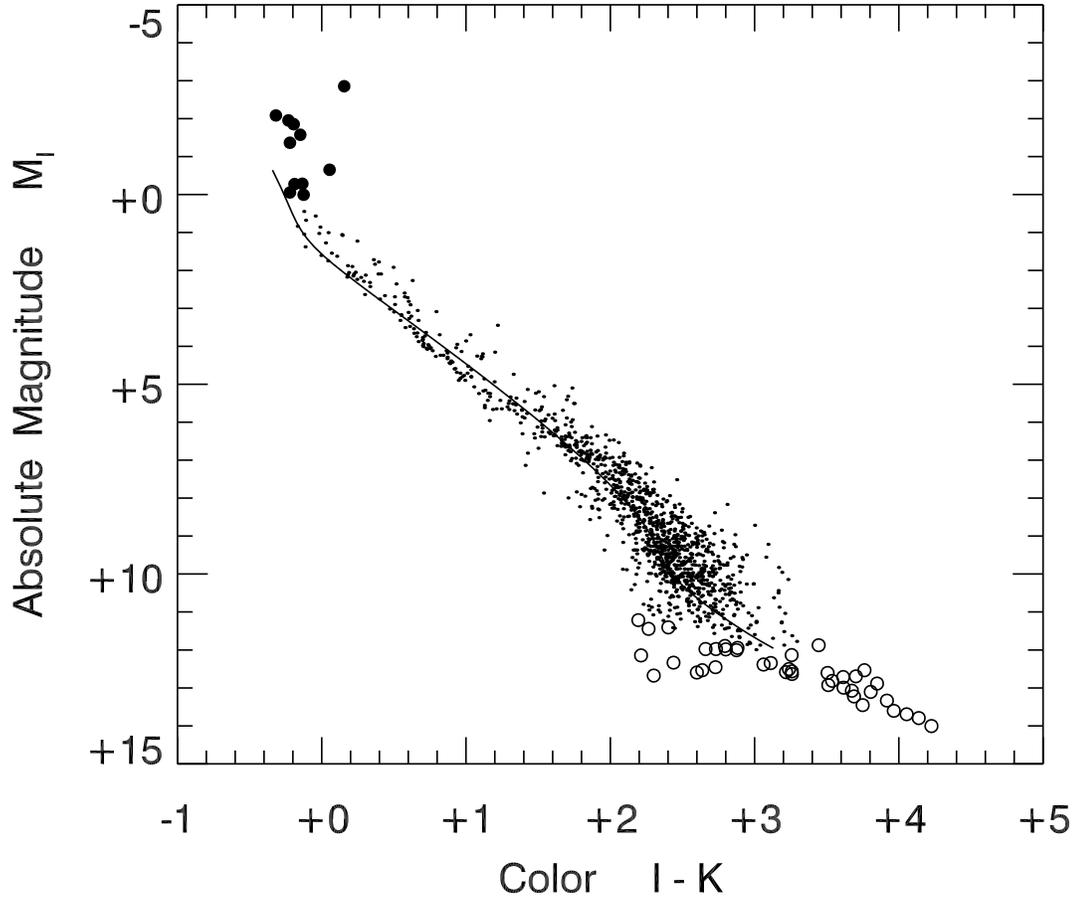}
\caption{Near-infrared color-magnitude diagram for the Pleiades. Small dots 
represent the 1245 stars in our sample. Open circles are the 41 likely 
sub-stellar objects which have been removed from the sample. Filled circles 
are the 11 brightest stars, which are likely post-main-sequence objects. The 
125 Myr isochrone for stars with masses between 0.08 and 4.0\,\,$\Msun$ is 
shown as the smooth, solid curve.}
\end{figure}

\clearpage

\begin{figure}
\plotone{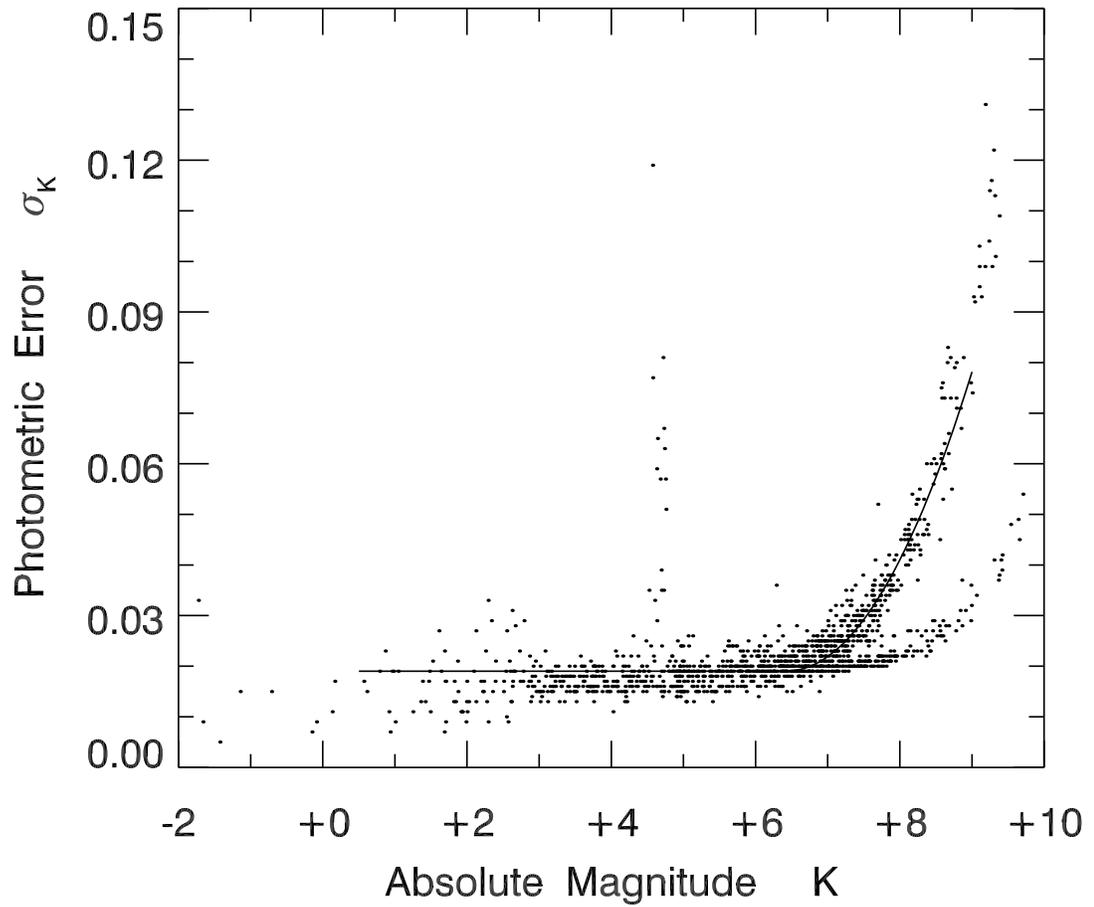}
\caption{Observational error in the $K$-band measurements as a function of
absolute magnitude for all 1417 stars in the catalog of \citet{sta07}. The 
smooth curve is the approximate fit used in our maximum likelihood analysis.}
\end{figure} 

\clearpage

\begin{figure}
\plotone{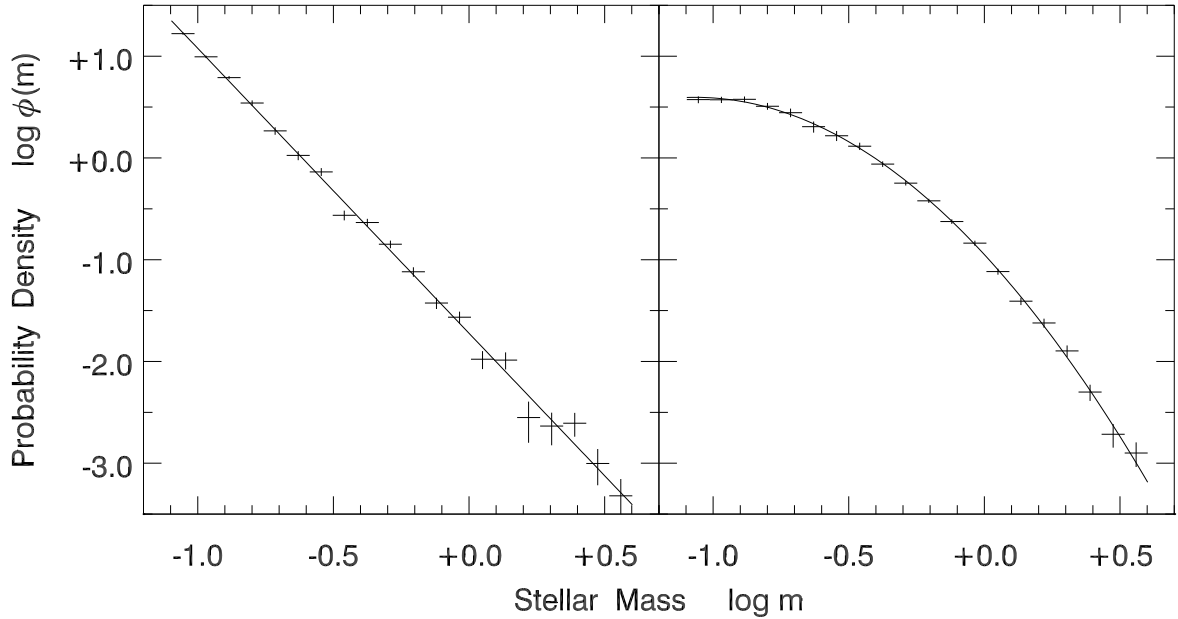}
\caption{Sample synthetic data results for the single star mass function
$\phi (m)$. In both panels, the smooth curve is the input function from which
the synthetic data were drawn. Shown also are the best-fit values, along with
errors, for our discrete representation of the function. In the left panel, 
the input $\phi (m)$  is a power law with slope -2.8; in the right panel, it 
is a log-normal function with peak \hbox{$m_\circ = 0.2$} and width 
\hbox{$\sigma_m\,=\,0.4$}.}
\end{figure} 

\clearpage
\begin{figure}
\plotone{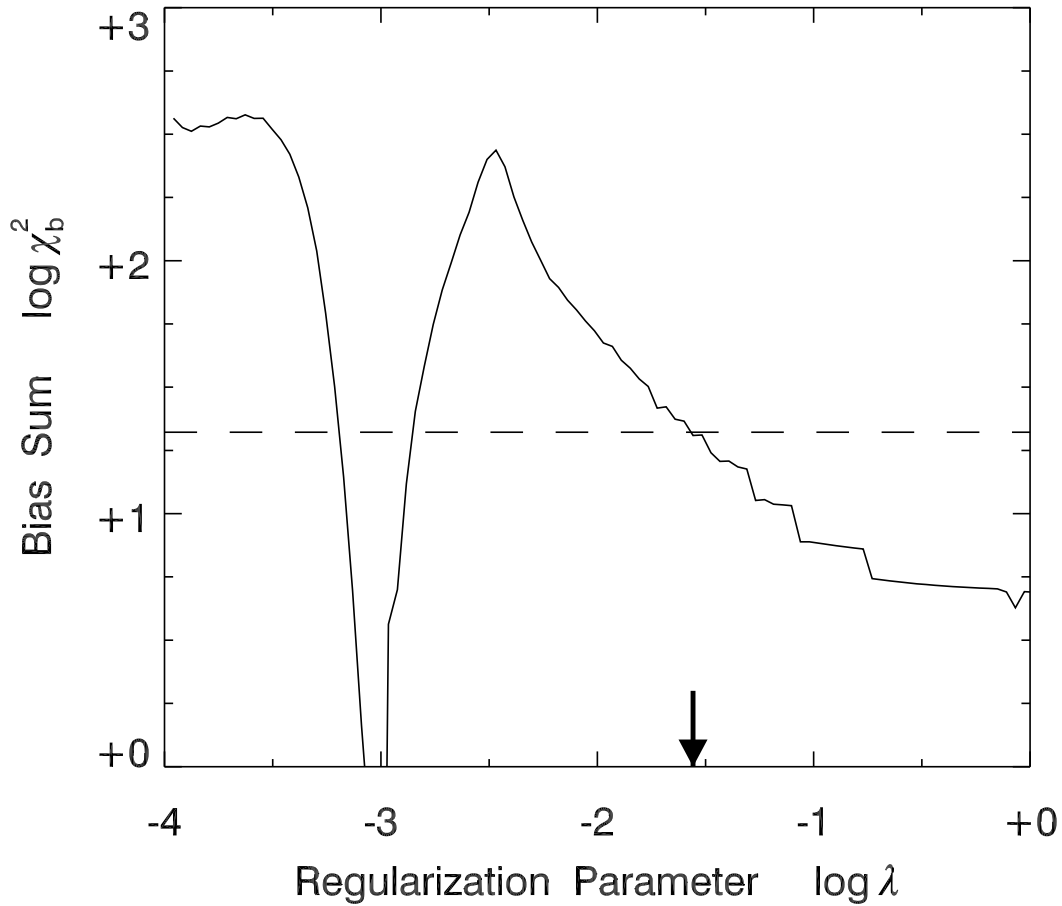}
\caption{Weighted sum of the biases as a function of the regularization 
parameter $\lambda$. The synthetic input here was a log-normal function with
1245 sources. The dashed, horizontal line is drawn at \hbox{$\chi_b^2 \,=\, N$},
where \hbox{$N\,=\, 21$} is the number of free parameters in our fitting. The
short, vertical arrow indicates the final value of $\lambda$ used for this
synthetic dataset.}
\end{figure} 

\clearpage
\begin{figure}
\plotone{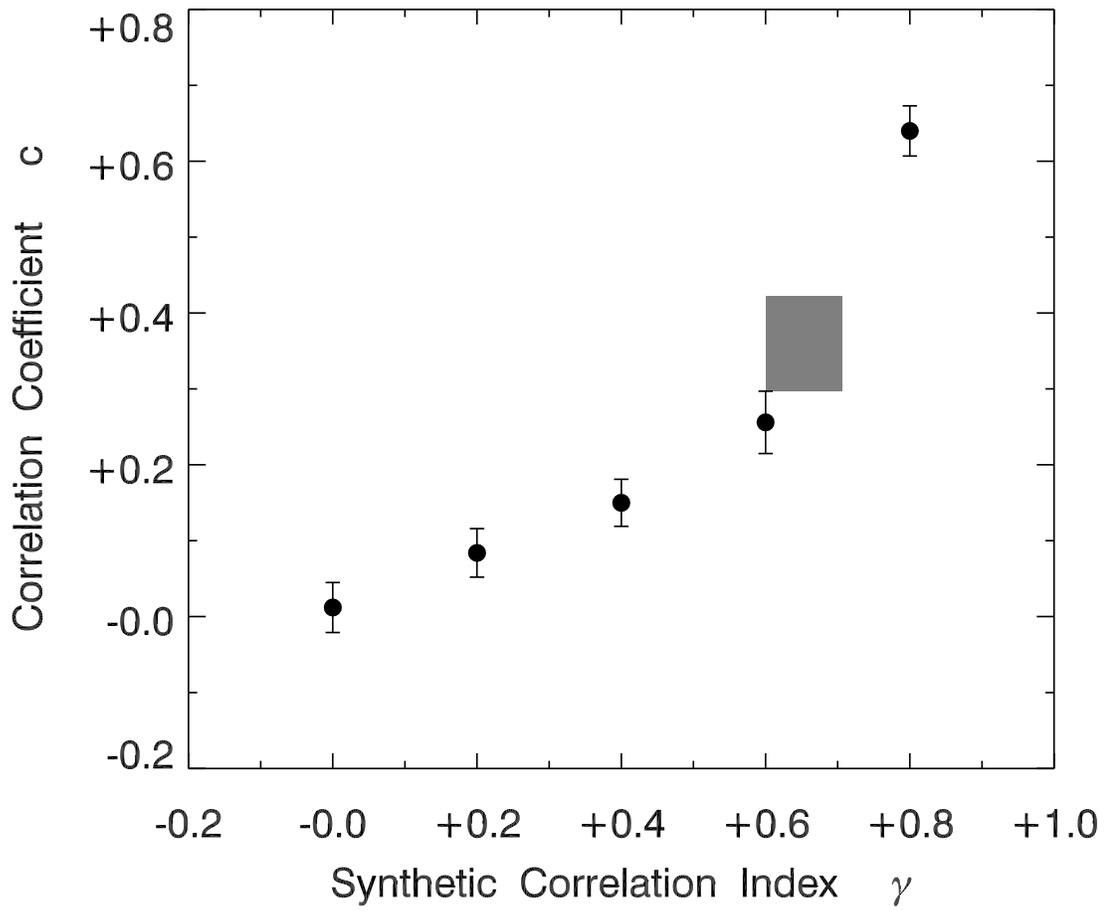}
\caption{Comparison of our fitted correlation coefficient $c$ with $\gamma$,
the imposed degree of correlation in the synthetic dataset. The gray area
indicates the region in which the Pleiades most likely falls.}
\end{figure} 

\clearpage
\begin{figure}
\plotone{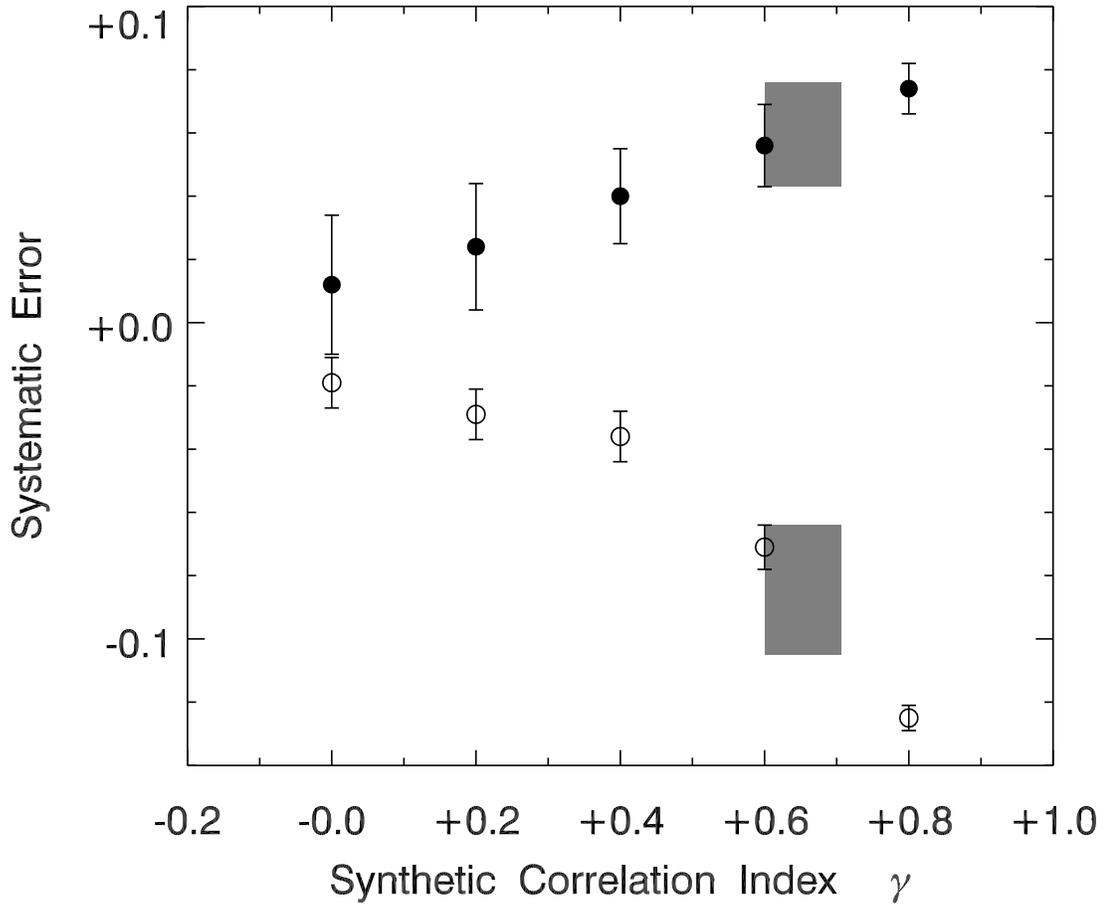}
\caption{Systematic errors in parameters of the log-normal fit to $\phi (m)$,
plotted as a function of the synthetic binary correlation $\gamma$. Filled
circles show $\Delta m_\circ$, the error in $m_\circ$. Open circles show
$\Delta\sigma_m$, the error in $\sigma_m$. The gray areas indicate the regions 
in which the Pleiades most likely falls.}
\end{figure} 

\clearpage
\begin{figure}
\plotone{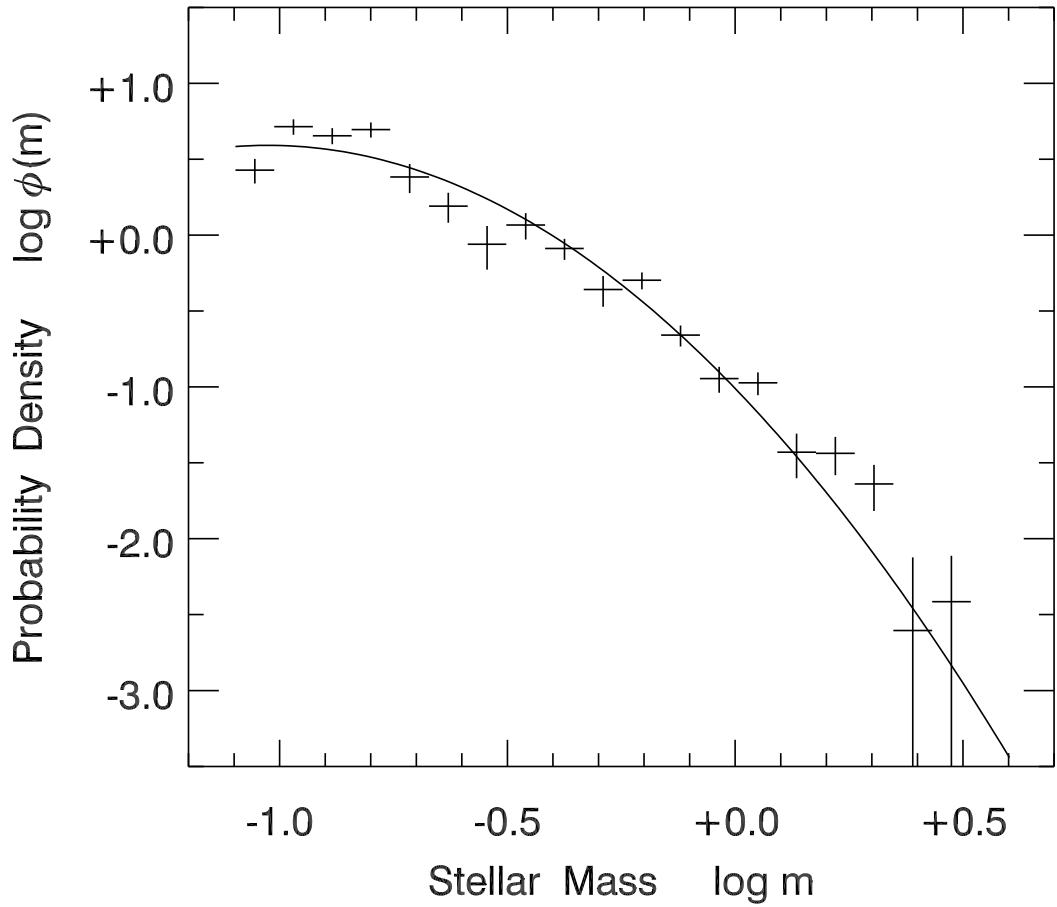}
\caption{Best-fit single star probability density $\phi (m)$ for the Pleiades.
Actual bin values $y_i/\Delta m_i$ are shown with associated errors. The smooth 
curve is a log-normal approximation to the results.}
\end{figure} 

\clearpage 
\begin{figure}
\plotone{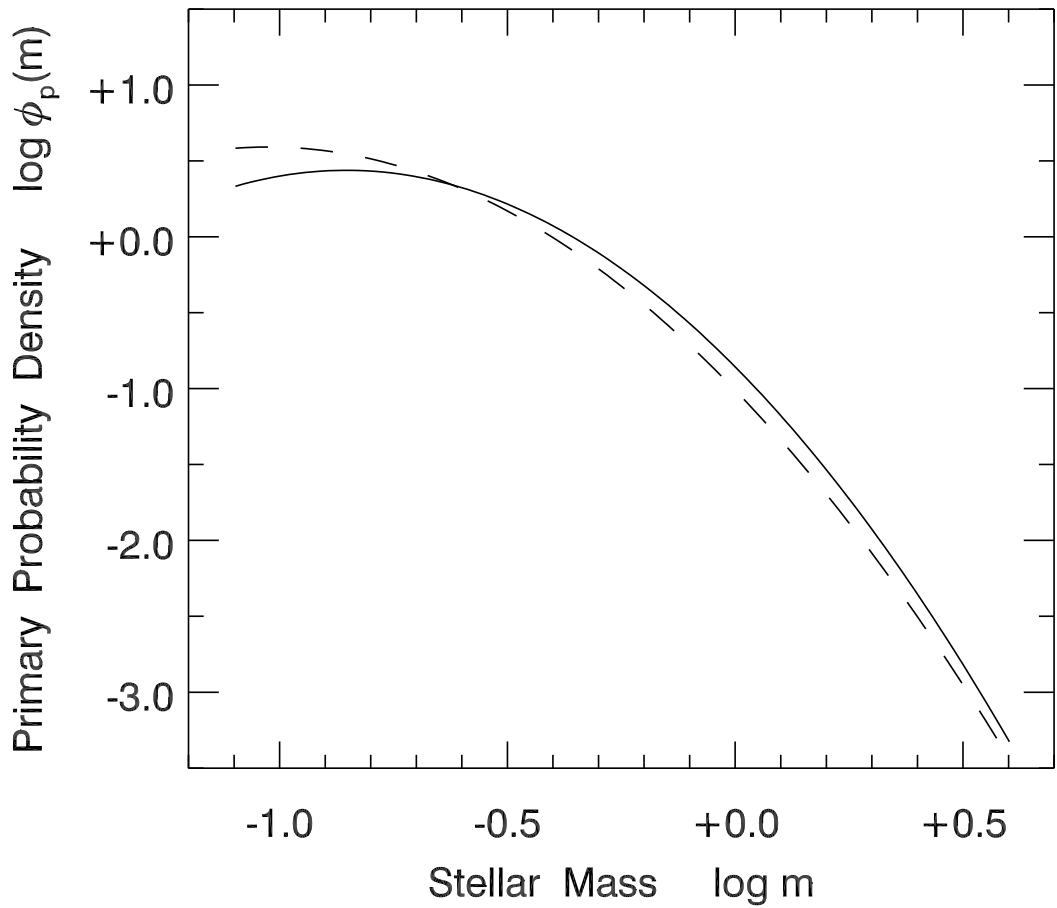}
\caption{Comparison of log-normal fits to the primary probability density
${\rm log}\,\phi_p (m_p)$ (solid curve) and the single star probability density
${\rm log}\,\phi (m)$ (dashed curve). The primary function peaks at larger
mass and has a smaller width. Note that $\phi_p (m)$ includes single stars.}
\end{figure} 

\clearpage
\begin{figure}
\plotone{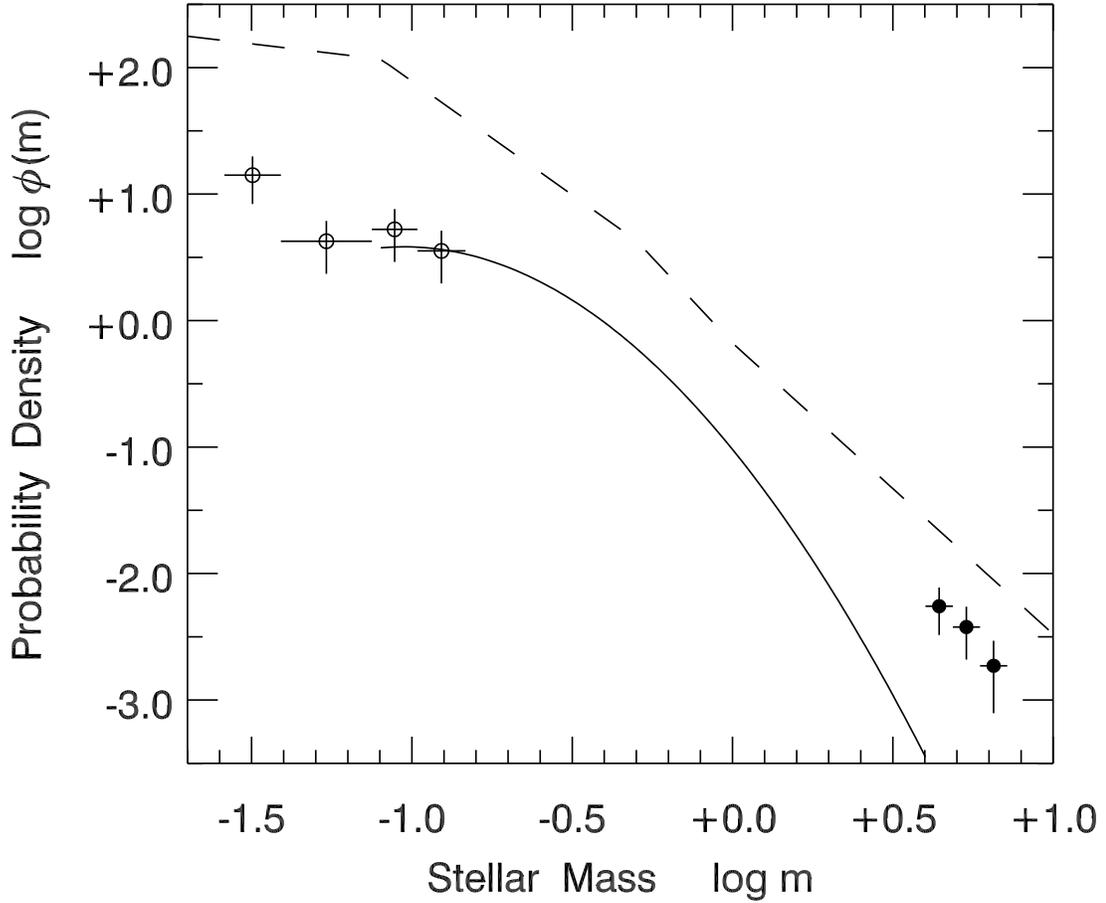}
\caption{Comparison of the Pleiades single star probability density to the 
field-star initial mass function of \citet{k01}, where the latter has been
shifted upward for clarity. Shown here is the log-normal approximation to 
${\rm log}\,\phi (m)$ (solid curve), augmented with the data of \citet{bi06} 
for low-mass members and brown dwarfs (open circles) and our 11 brightest, 
post-main-sequence stars (filled circles).}  
\end{figure}  

\clearpage 
\begin{figure}
\plotone{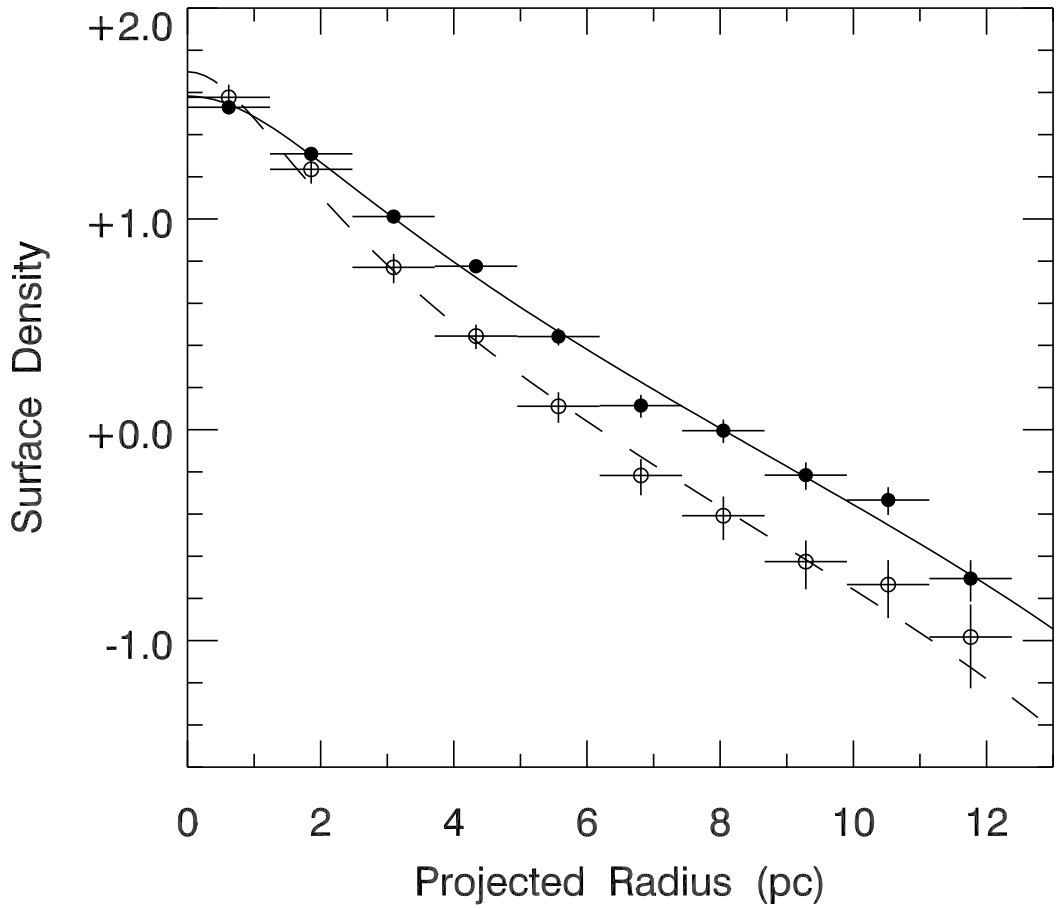}
\caption{Surface density distribution in the Pleiades. The filled circles
represent the surface number density (${\rm pc}^{-2}$), displayed on a
logarithmic scale. Open circles are the mass density, in 
\hbox{$\Msun\,\,{\rm pc}^{-2}$}. The solid and dashed smooth curves are King
model fits.}
\end{figure} 

\clearpage  
\begin{figure}
\plotone{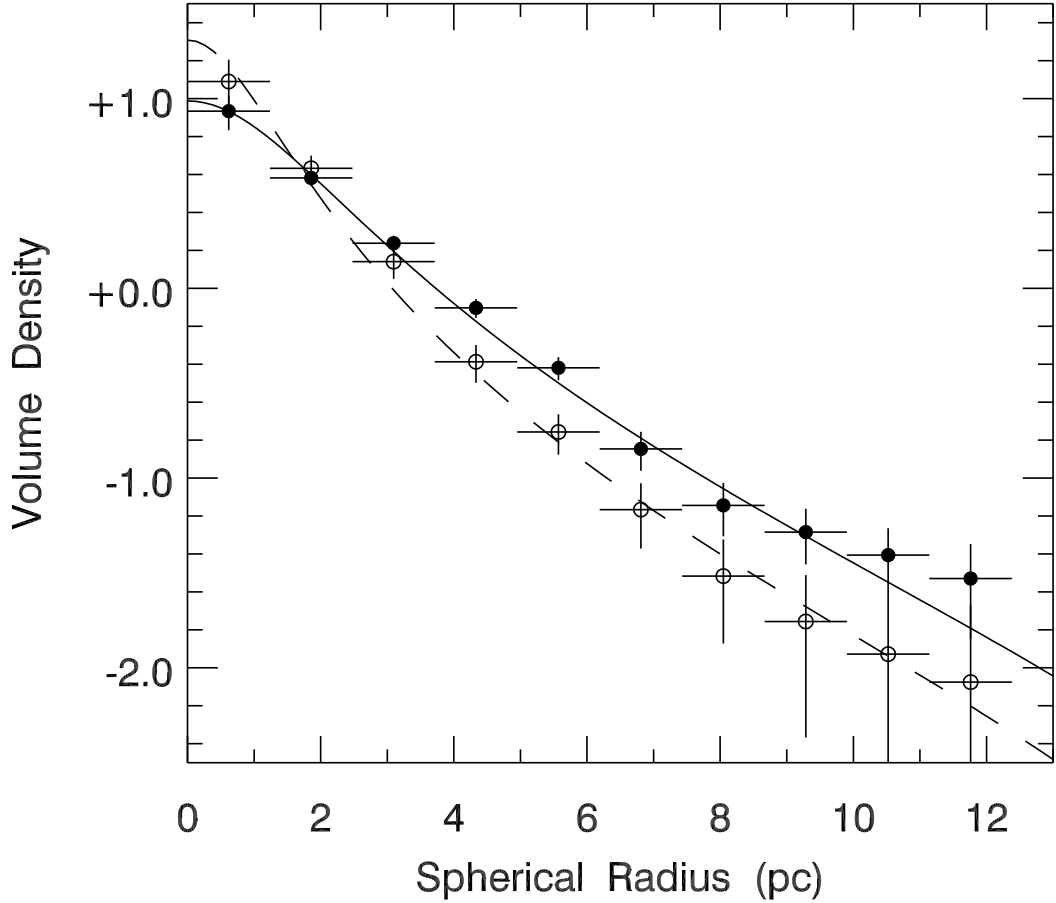}
\caption{Volume density profiles. The filled circles
represent the number density (${\rm pc}^{-3}$), again displayed 
logarithmically. Open circles are the mass density, in  
\hbox{$\Msun\,\,{\rm pc}^{-3}$}. The solid and dashed smooth curves are the 
same King model fits as in Figure~10, but now deprojected into 
three-dimensional space.} 
\end{figure}
 
\clearpage  
\begin{figure}
\plotone{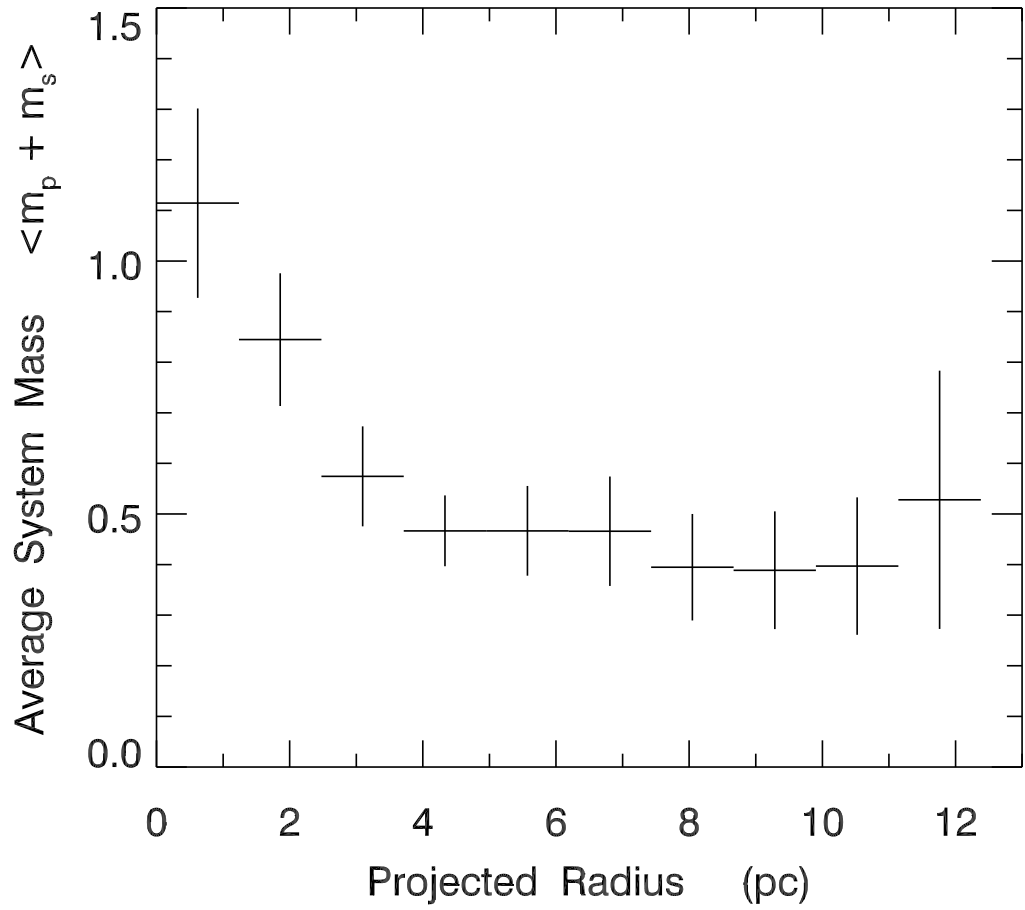}
\caption{Average system mass (primaries plus secondaries) as a function of
projected cluster radius. The bins here have constant radial width.}
\end{figure}
 
\clearpage
\begin{figure}
\plotone{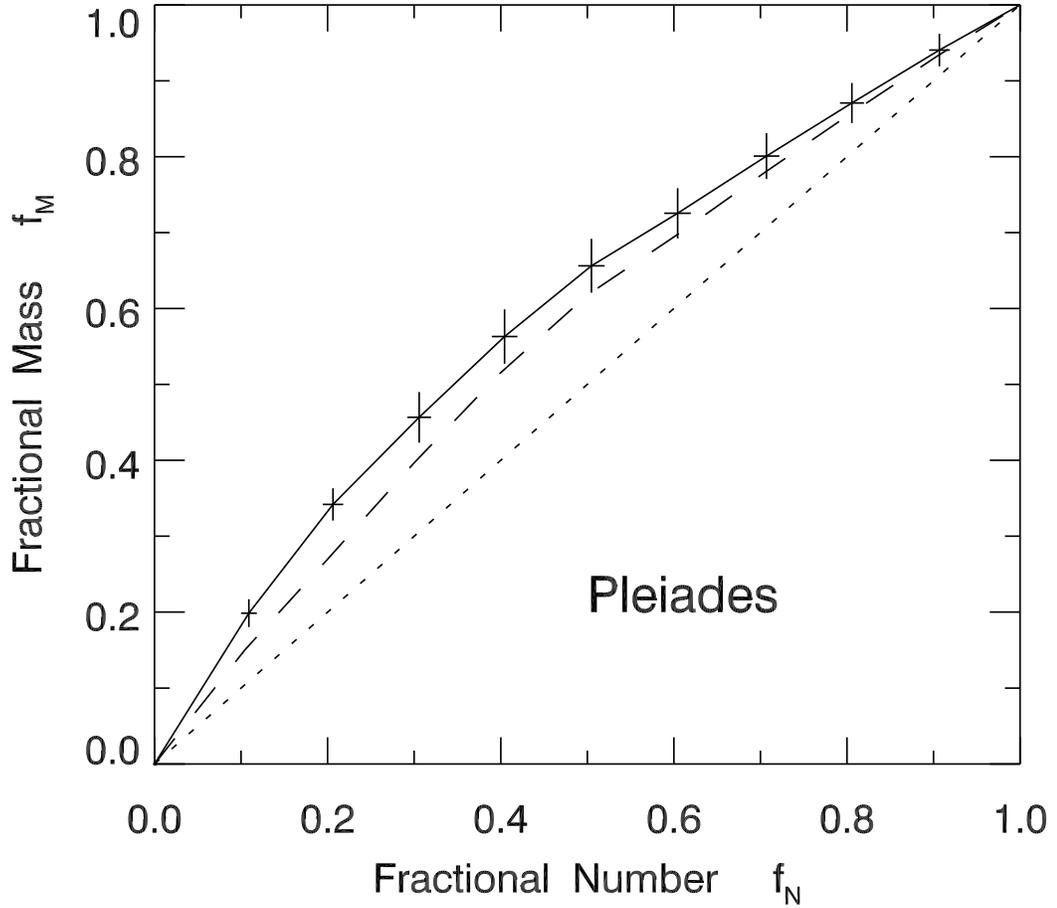}
\caption{Fractional mass versus fractional number for the Pleiades. The data
points and error bars, along with the solid curve, utilize all sources in the
catalog of \citet{sta07}. The dashed curve shows the result when the 11 
brightest stars are removed. In both cases, the radial bins contain roughly
equal numbers of stars. Finally, the dotted diagonal is the hypothetical
result for no mass segregation.} 
\end{figure}

\clearpage  
\begin{figure}
\plotone{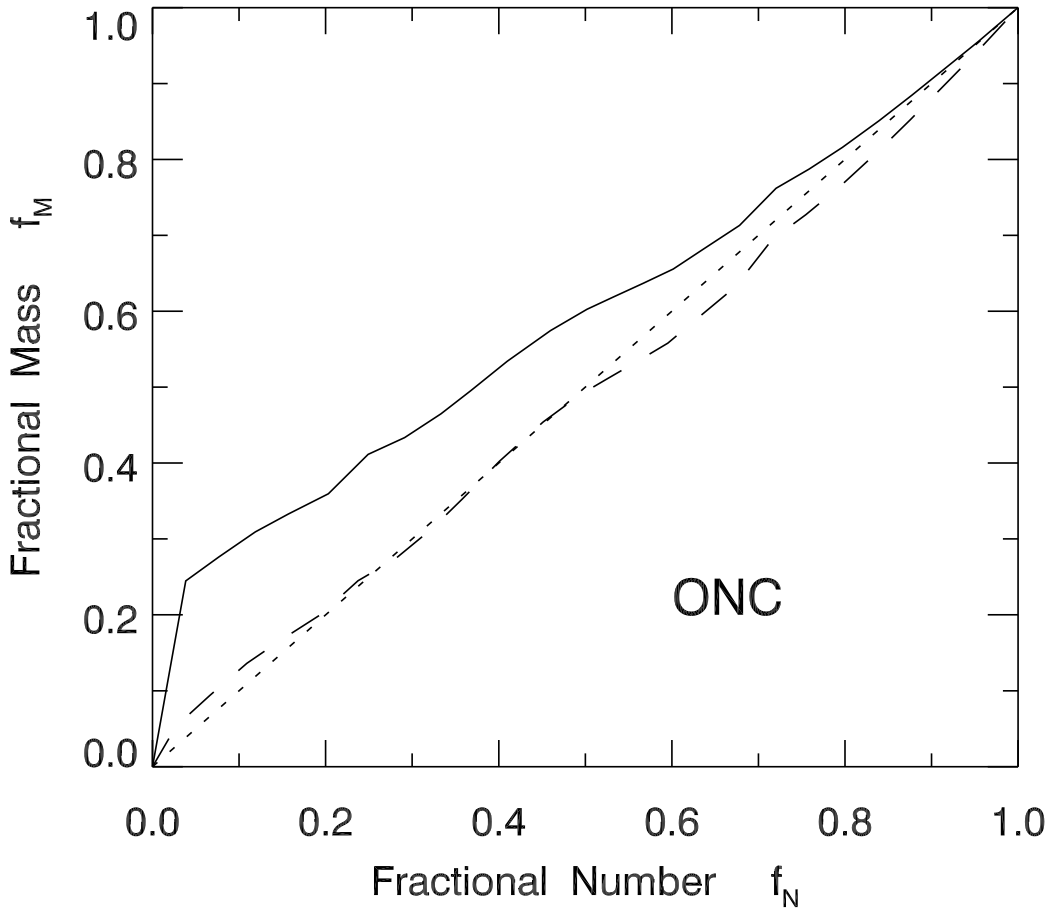}
\caption{Fractional mass versus fractional number for the ONC. The solid curve
was computed using all the ONC sources analyzed by \citet{hs06}. The dashed
curve shows the result if only the 4 Trapezium stars are removed. As in 
Figure~13, the radial bins contain roughly equal numbers of stars. The dotted
diagonal again shows the hypthetical condition of no mass segregation.}
\end{figure} 


\end{document}